\documentclass[manuscript,sigconf,dvipsnames]{acmart}

\acmVolume{0}
\acmNumber{0}
\acmArticle{0}
\acmMonth{0}

\copyrightyear{2022} 
\acmYear{2022} 
\setcopyright{rightsretained} 
\acmConference[RecSys '22]{Sixteenth ACM Conference on Recommender Systems}{September 18--23, 2022}{Seattle, WA, USA}
\acmBooktitle{Sixteenth ACM Conference on Recommender Systems (RecSys '22), September 18--23, 2022, Seattle, WA, USA}
\acmDOI{10.1145/3523227.3546765}
\acmISBN{978-1-4503-9278-5/22/09}

\usepackage{enumitem}
\usepackage{multirow}
\usepackage{amsmath}

\graphicspath{{figure/}{figures/}}

\usepackage{pgfplots}
\pgfplotsset{compat=newest}
\usetikzlibrary{patterns}

\usepackage{subcaption}
\DeclareCaptionFont{tiny}{\tiny} 
\newcommand{\algorithmautorefname}{Algorithm}

\usepackage{algorithm}
\usepackage[noend]{algorithmic}

\newcommand{\algorithmicprocedure}{\textbf{procedure}}
\newcommand{\algorithmicendprocedure}{\algorithmicend\ \algorithmicprocedure}
\makeatletter
\newcommand\PROCEDURE[3][default]{%
  \ALC@it
  \algorithmicprocedure\ \textsc{#2}(#3)%
  \ALC@com{#1}%
  \begin{ALC@prc}%
}
\newcommand\ENDPROCEDURE{%
  \end{ALC@prc}%
  \ifthenelse{\boolean{ALC@noend}}{}{%
    \ALC@it\algorithmicendprocedure
  }%
}
\newenvironment{ALC@prc}{\begin{ALC@g}}{\end{ALC@g}}
\makeatother

\newcommand{\pseudocode}[3]{%
    \begin{minipage}{\hsize}%
        \renewcommand\figurename{Algorithm}%
        \setcounter{tmp}{\value{figure}}%
        \setcounter{figure}{\value{algorithm}}%
        \stepcounter{algorithm}%
        \smaller\smaller%
        \noindent\rule{\hsize}{0.8pt}%
        \captionsetup{labelfont={small,bf},textfont={small},labelsep=space,justification=raggedright,singlelinecheck=false}%
        \vspace{-1.65em}\caption{#2}\label{#1}%
        \vspace{-0.7em}\noindent\rule{\hsize}{0.4pt}%
        \vspace{-1.5em}\begin{algorithmic}[1]#3\end{algorithmic}%
        \vspace{-0.9em}\noindent\rule{\hsize}{0.4pt}%
        \setcounter{figure}{\value{tmp}}%
    \end{minipage}%
}
\newcommand{\pseudoref}[1]{\hyperref[#1]{\algorithmautorefname} \ref{#1}}

\colorlet{LineColorA}{RoyalBlue}
\colorlet{ShadeColorA}{RoyalBlue!20!White}

\colorlet{LineColorB}{Orange}
\colorlet{ShadeColorB}{Orange!20!White}

\colorlet{LineColorC}{Gray}
\colorlet{ShadeColorC}{Gray!20!White}

\colorlet{LineColorC}{PineGreen}
\colorlet{ShadeColorC}{PineGreen!20!White}

\colorlet{LineColorD}{Dandelion}
\colorlet{ShadeColorD}{Dandelion!20!White}

\colorlet{LineColorE}{Violet}
\colorlet{ShadeColorE}{Violet!20!White}

\colorlet{LineColorF}{Salmon}
\colorlet{ShadeColorF}{Salmon!20!White}

\begin{document}

\title[HugeCTR Inference Parameter Server]{%
A GPU-specialized Inference Parameter Server for Large-Scale Deep Recommendation Models%
}

\author{Yingcan Wei}
\email{yingcanw@nvidia.com}
\orcid{0000-0002-5093-7382}
\affiliation{%
  \institution{NVIDIA}
  \city{Shanghai}
  \country{China}
}

\author{Matthias Langer}
\email{mlanger@nvidia.com}
\orcid{0000-0003-1776-8000}
\affiliation{%
  \institution{NVIDIA}
  \city{Shanghai}
  \country{China}
}

\author{Fan Yu}
\email{fayu@nvidia.com}
\orcid{0000-0001-8454-3923}
\affiliation{%
  \institution{NVIDIA}
  \city{Shanghai}
  \country{China}
}

\author{Minseok Lee}
\email{minseokl@nvidia.com}
\orcid{0000-0002-8367-1939}
\affiliation{%
  \institution{NVIDIA}
  \city{Seoul}
  \country{South Korea}
}

\author{Kingsley Liu}
\email{kingsleyl@nvidia.com}
\orcid{0000-0002-4293-4827}
\affiliation{%
  \institution{NVIDIA}
  \city{Shanghai}
  \country{China}
}

\author{Jerry Shi}
\email{jershi@nvidia.com}
\orcid{0000-0003-1446-0326}
\affiliation{%
  \institution{NVIDIA}
  \city{Shanghai}
  \country{China}
}

\author{Joey Wang}
\email{zehuanw@nvidia.com}
\orcid{0000-0002-1072-2651}
\affiliation{%
  \institution{NVIDIA}
  \city{Beijing}
  \country{China}
}

\renewcommand{\shortauthors}{Y. Wei \emph{et al.}}

\begin{abstract}
    Recommendation systems are of crucial importance for a variety of modern apps and web services, such as news feeds, social networks, e-commerce, search, \emph{etc}. To achieve peak prediction accuracy, modern recommendation models combine deep learning with terabyte-scale embedding tables to obtain a fine-grained representation of the underlying data. Traditional inference serving architectures require deploying the whole model to standalone servers, which is infeasible at such massive scale.

    In this paper, we provide insights into the intriguing and challenging inference domain of online recommendation systems. We propose the HugeCTR Hierarchical Parameter Server (HPS), an industry-leading distributed recommendation inference framework, that combines a high-performance GPU embedding cache with an hierarchical storage architecture, to realize low-latency retrieval of embeddings for online model inference tasks. Among other things, HPS features (1) a redundant hierarchical storage system, (2) a novel high-bandwidth cache to accelerate parallel embedding lookup on NVIDIA GPUs, (3) online training support and (4) light-weight APIs for easy integration into existing large-scale recommendation workflows. To demonstrate its capabilities, we conduct extensive studies using both synthetically engineered and public datasets. We show that our HPS can dramatically reduce end-to-end inference latency, achieving 5\textasciitilde62x speedup (depending on the batch size) over CPU baseline implementations for popular recommendation models. Through multi-GPU concurrent deployment, the HPS can also greatly increase the inference QPS.
\end{abstract}



\maketitle
\setlength{\fboxsep}{0pt}%
\newcounter{tmp}%

\section{Introduction}\label{s:intro}

\begin{figure*}[tb]
    \begin{minipage}[b]{0.475\hsize}%
        \centering%
        \includegraphics[width=0.8\hsize]{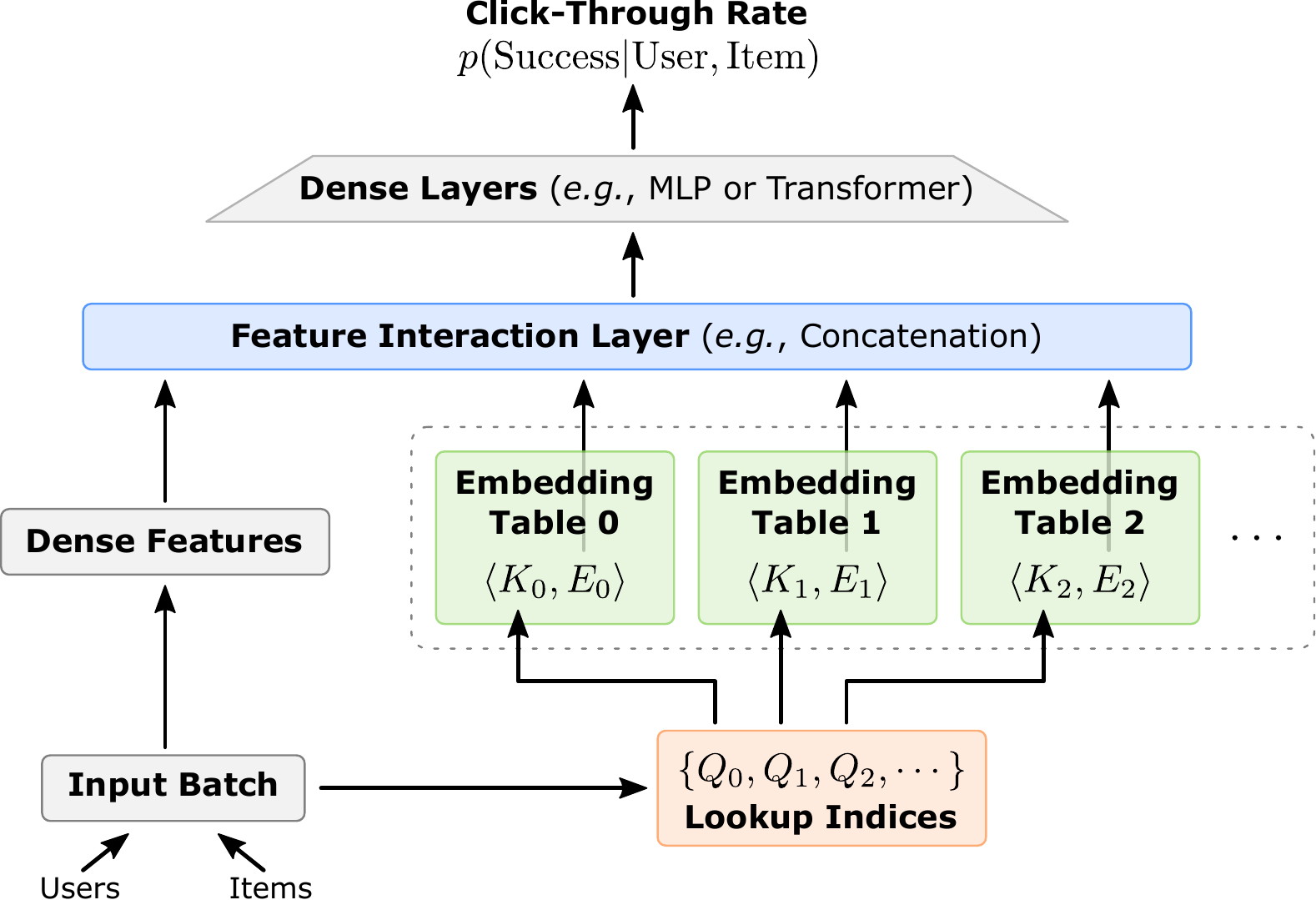}%
        \caption{A typical deep recommendation model (DLRM).}%
        \label{f:simple-ctr-model}%
    \end{minipage}%
    \hfill%
    \begin{minipage}[b]{0.475\hsize}%
        \centering%
        \includegraphics[width=0.9\hsize]{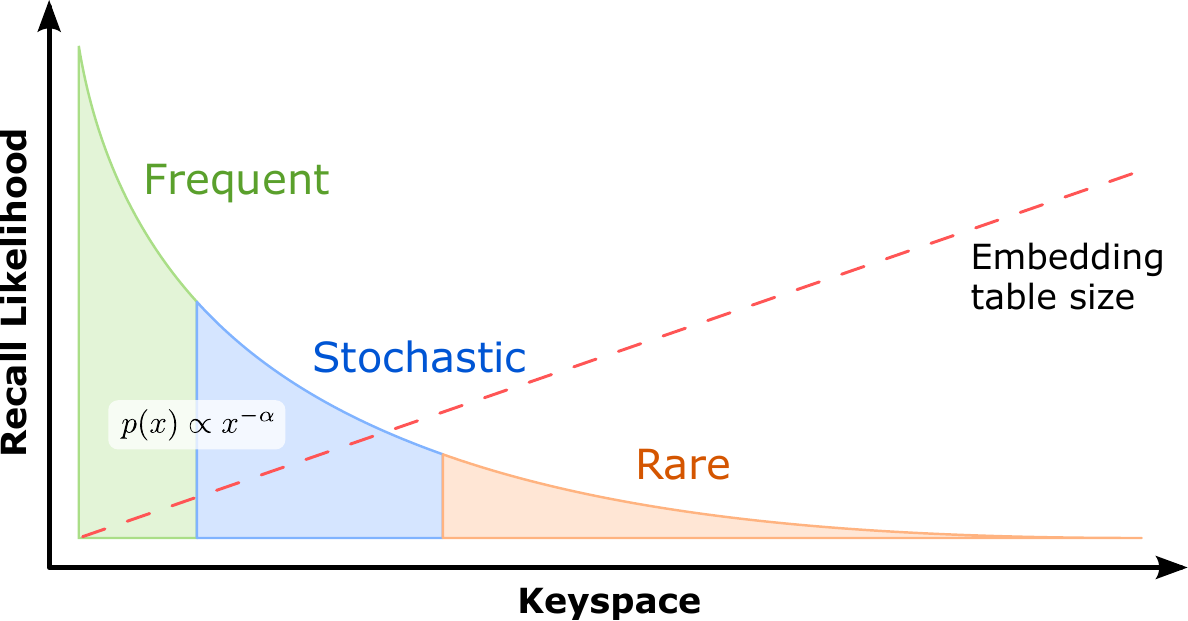}%
        \caption{Key distribution if recall statistics approximately follow a power law distribution.}%
        \label{f:power-law}%
    \end{minipage}%
\end{figure*}

Recommendation Systems (RS) are used in various apps and online services, such as news feeds, e-commerce, social networks, search, \emph{etc}. To provide accurate predictions, state-of-the-art algorithms rely on embedding-based deep learning models. \autoref{f:simple-ctr-model} illustrates the typical architecture of a deep recommendation model (DLRM). The input consists of dense features (\emph{e.g.}, age, price, \emph{etc.}) and sparse features (\emph{e.g.}, user ID, category ID, \emph{etc.}). The sparse features are transformed into dense embedding vectors through lookup in an embedding table, so that the result from combining these with the dense features can be fed through some densely connected deep learning model (\emph{e.g.}, a MLP, transformer, \emph{etc}. \cite{vaswani2017attention,sun2019bert4rec}) to predict the Click-Through Rate (CTR).

Embeddings can consume a significant portion of the memory capacity in a data center. Often, a significant amount of time is spent to retrieve these embeddings from a centralized parameter server, which adds latency that delays downstream computations. Unlike in throughput-oriented training systems \cite{DeepRecSys,ScaleFreeCTR,GeePS,Clipper,Angel,Distributed_Hierarchical,FlexPS,Parameter_Hub,Fast_Distributed_Training}, online inference systems are tightly constrained by latency requirements \cite{RecSSD}. Thus, the embedding lookup speed is essential for deep recommendation model inference performance.

During inference, each mini-batch of data usually references tens of thousands of embeddings. Realizing the exhaustive search of each embedding by its key requires the parameter server to walk certain internal data structures. The lookup of individual embeddings from an embedding table is usually independent, and, thus, easily parallelizable. At the same time, modern GPU architectures allow scheduling thousands of threads to run concurrently, and their memory subsystems adopt special memory technology that provides higher bandwidth and throughput than equivalent CPU memories \cite{nvidia2020ampere-arch}. These features could make GPU architectures ideal for processing embedding vector lookup workloads.

\textbf{Challenges.} The size of embedding tables used in state-of-the-art recommendation models can be vast, often ranging from tens of giga- to several terabytes, which is well beyond the memory complement of most GPUs. Furthermore, batch sizes during online inference are usually too small to efficiently utilize the massively parallel-processing-optimized computational resources of just a single GPU. Hence, embedding lookup workloads require large amounts of GPU memory, but only few computational resources. This imbalance of requirements significantly deviates from the available hardware, and diminishes GPUs' attractiveness for use in inference systems. Therefore, most existing solutions decouple the embedding lookup operation from the dense computations (\emph{i.e.}, the remainder of the model), which are executed in the GPU, and move it to the CPU \cite{Capacity-Driven}. Thereby, they forfeit the memory bandwidth advantages of GPUs, while the CPU and the communication bandwidth between the CPU and the GPU becomes the primary bottleneck. As a result, the disproportionate processing capabilities of GPUs sit mostly idle in such setups (=resource waste).

\textbf{Approach.} It is usually not possible to retain all embedding tables entirely in GPU memory. However, empirical evidence for real-world recommendation datasets suggests that embedding key access during inference for CTR and other recommendation tasks often exhibits strong locality, and approximately follows the power law distribution \cite{ScaleFreeCTR,GeePS,Clipper,Angel}. Hence, a significant proportion of the embedding keys per mini-batch reference only a small set of hot embeddings. Caching such hot embeddings in the GPU memory, where the remainder of the model is processed, makes partial GPU-accelerated embedding lookup possible. Based on these observations, we have built an inference framework, namely the HugeCTR Hierarchical Parameter Server (HPS), to take advantage of GPU resources, without being constrained to GPU memory limitations. In particular, HPS introduces a GPU embedding cache data structure that tries to retain hot embeddings within the GPU memory. The cache is complimented by a parameter server that keeps a full copy of all embedding tables. Our contributions can be summarized as follows:
\begin{itemize}
    \item \emph{Hierarchical database architecture} that allows utilizing cluster memory resources, and provides an asynchronous update mechanism to maintain a high GPU embedding cache hit rate during online inference.
    \item \emph{High-performance dynamic GPU embedding cache} that maximizes throughput by tracking and caching frequently occurring embeddings in high-throughput GPU memory, while overlapping the host/device transfers.
    \item \emph{Online model update mechanism} for distributed inference deployments (\emph{i.e.}, real-time updates).
    \item \emph{Customizable HPS backend} that provides concurrent model execution, hybrid model deployment, and ensemble model pipeline services for NVIDIA Triton GPU inference server \cite{nvidia2022hugectr-backend}.
\end{itemize}
This paper is structured as follows. In \autoref{s:problem-formulation}, we provide a fundamental discussion of core concepts that underpin our approach. Then, we subsequently introduce and discuss the individual components of the HPS and how they interact in Sections \ref{s:hps}\textasciitilde\ref{s:hps-db}. In \autoref{s:update}, we discuss how our HPS realizes real-time model updates. Eventually, we conduct an experimental study to evaluate the performance of the HPS in \autoref{s:performance-evaluation}, and provide concluding remarks in \autoref{s:conclusion}.

\section{Background}\label{s:problem-formulation}

\subsection{Embedding Tables}
Current mainstream algorithms in advertising, recommendation and search adopt model structures that combine embedding tables with a deep neural network to form a deep learning recommendation model (DLRM) \cite{dlrm}. At the foundation of such models are embeddings $e$, which represent learned numeric representations of user or item features as dense vectors that are aligned in some $d$-dimensional space ($e \in \mathbb{R}^d$). We let $E_j = \{ e_j^0, e_j^1, \cdots, e_j^n \}$ denote some discrete subset of the embeddings for some feature $j$. For easy access within the model, we organize these embeddings as embedding feature tables of the form
\begin{equation}
    T_j = \langle K_j, E_j \rangle = \lbrace \langle k_j^0, e_j^0 \rangle, \langle k_j^1, e_j^1 \rangle, \cdots, \langle k_j^n, e_j^n \rangle \rbrace ~,
\end{equation}
that consist of tuples $\langle k_j^i, e_j^i \rangle$, where $k_j^i$ is a key that identifies and references the $i$-th embedding table entry $e_j^i$. The key space $K_j = \{ k_j^0, k_j^1, \cdots, k_j^n \}$ is discrete, and $(\forall k_j^i, k_j^z \in K_j)(i \neq z \rightarrow k_j^i \ne k_j^z)$ is implied. The value of each key depends on the underlying data or task. Usually, the key space is sparsely populated.

To evaluate a DLRM for CTR (\emph{cf.} \autoref{f:simple-ctr-model}), the driver application must first select the entries from the embedding table that are relevant to make the prediction. This is simply done by looking up\footnote{Direct key lookup is the predominantly used method. For complex models other methods to determine the query keys $Q$ may exist.} the keys from the query key subset $Q_j$ for each embedding feature table (\emph{i.e.}, $Q = \{ Q_0 \subseteq K_0, Q_1 \subseteq K_1, \cdots \}$). Thus, $Q_j = \{ q_j^0, q_j^1, \cdots, q_j^m \}$ denotes a query for looking up the $m$ corresponding embeddings entries from $T_j$. The corresponding result set is $R_{Q_j} = \{ q_j^0 \mapsto e_j^0, q_j^1 \mapsto e_j^1, \cdots, q_j^m \mapsto e_j^m \}$. Accelerating the retrieval of such result sets at scale is our primary objective.

\subsection{Deduplication and skewness.}\label{s:power-law}
To avoid unnecessary double-lookups if the same embedding table entries are required multiple times, HugeCTR always applies a deduplication operator prior to executing any subsequent steps (\emph{i.e.}, $Q^* = \text{\textsc{dedup}}(Q)$). This is particularly important for mini-batch processing, where $Q$ is the concatenation of many input samples. Naturally, deduplication becomes more effective if the skewness of the query distribution $\mathcal{Q}$ increases.

Understanding and utilizing skewness properties of the dataset is peril achieving peak efficiency. Many real-world recommendation datasets (\emph{e.g.}, Criteo \cite{criteo2014dataset}) exhibit a power law distribution \cite{clauset2009power-law-dist}. That is, certain subsets of keys are referenced more frequently than others, such that sampling $q_j \sim \mathcal{Q}_j$ eventually approximates $p(x) \propto x^{-\alpha}$. \autoref{f:power-law} depicts a scenario where the embedding key recall statistics approximate a power law distribution. The key space can be divided into three categories: (1) \textbf{Frequent} embeddings effectively appear in every batch. They represent a significant fraction of the recall/update requests. The frequent set is usually small. Even for large embedding corpora, only up to a few thousand embeddings appear that regularly. (2) \textbf{Stochastic} embeddings appear every few batches (\emph{i.e.}, somewhat regularly over time). (3) \textbf{Rare} embeddings are at the far end of the spectrum. They appear rather infrequently in queries. 

Because requests repeatedly reference \emph{frequent} and \emph{stochastic} embeddings, applying efficient caching methods to them improves the overall system performance the most. Our HPS design (see \autoref{s:hps}) builds up on this observation.

Such category assignments of embeddings are absolutely determined if the query dataset is fixed. When training HugeCTR models, we take advantage of this to achieve world-class model convergence rates \cite{kanter2021mlperf-1-1-results,farrell2021mlperf}. During online inference, the recall statistics depend on the actually incoming user requests. These cannot be preempted. Due to sudden events, changing trends or fashion, the category assignment of individual embeddings can vary over time. For most recommendation tasks, the runtime statistics are in a constant flux. Thus, inference systems must be adaptive.

\subsection{GPU-accelerated Inference Architecture}\label{s:inf-arch}
Parameter servers for ML inference workloads mostly rely on database operations that are trivially parallelizable with GPUs \cite{yuan2013gpu-query-processing,subramanian2021gpu-db-operators-benchmark,rosenfeld2022cpu-gpu-query-processing-survey}. Applications that require fast response times, \emph{e.g.}, online transaction processing (OLTP), often benefit greatly from GPU acceleration \cite{arefyeva2018cpu-gpu-db-survey}. However, GPU memory constraints pose a tough challenge. To achieve scalability, many existing GPU-accelerated database systems, as well as our approach, implement a hierarchical storage architecture that extends the available GPU memory with other storage resources. Because external memory resources cannot be accessed as efficiently as native GPU memory \cite{nvidia2020ampere-arch}, the data exchange performance with the host system is emphasized in such systems \cite{mittal2015cpu-gpu-heterogeneous-survey}. To achieve peak performance, overlapped query processing must be used in conjunction with efficient communication patterns and data placement strategies that are actively refined at runtime \cite{arefyeva2018cpu-gpu-db-survey,bennun2019dist-deep-learning,langer2020ddls}.

Constructing a parameter server for a machine learning platform poses many challenges \cite{bennun2019dist-deep-learning,langer2020ddls,DeepRecSys,ScaleFreeCTR,GeePS,Clipper,Angel,Distributed_Hierarchical,FlexPS,Parameter_Hub,Fast_Distributed_Training}. When designing mixed GPU/CPU-based architectures for inference production environments, at least two major bottlenecks must be overcome:
(1) \textbf{High latency due to DRAM bandwidth limitations} when communicating between CPU and GPU \cite{RecSSD,jiang2021fleet-rec}.
(2) \textbf{Deployment latency} due to growing model size and complexity induced by online training, because fast-paced incremental model updates pose a great challenge with respect to data consistency and bandwidth. To address these bottlenecks, our HPS is specifically tailored for deployment as an inference parameter server for large-scale recommendation models on GPUs. It handles the data synchronization and communication to share model parameters (embedding tables) across different inference nodes \cite{nvidia2022hps}, and performs various optimizations to improve GPU utilization during parallel multi-model/multi-GPU inference, including the organization of the distributed embedding table into partitions \cite{nvidia2022hps-dist-deployment}, GPU-friendly caching \cite{nvidia2022hps-gpu-cache}, and an asynchronous data movement mechanism \cite{nvidia2022hps-async-data-insert}.

\section{Hierarchical Parameter Server}\label{s:hps}

Our Hierarchical Parameter Server (HPS) allows HugeCTR to use models with huge embedding tables for inference. This is achieved through extending the embedding storage space beyond the constraints of GPUs using CPU memory resources from across the cluster. The design target of the HPS is to address the three challenges that traditional CPU parameter server approaches typically suffer from most:
\begin{enumerate}
    \item \textbf{Downloading/streaming of model parameters} from the centrally maintained embedding table partitions in CPU memory to the model instances on individual GPU compute devices. This issue is magnified if the embedding table cannot be fit entirely into the GPU memory. HPS greatly alleviates this problem through a GPU caching mechanism that takes advantage of the locality of the data distribution.
    \item \textbf{Increased deployment cost} caused by high-availability requirements of inference platforms and bandwidth limitations. By jointly organizing and using the distributed CPU memories of the inference cluster, the HPS saves resources and realizes immediate online model updating (\emph{i.e.}, training to inference updates).
    \item \textbf{Parameter update and refresh} between the GPU cache and the parameter server. This is particularly challenging if only a part of the model is loaded into GPU memory, so that parameters are missed on the GPU during lookup. HPS handles additional parameter exchanges between the CPU and GPU using an asynchronous insertion and refreshing mechanism to maintain parameter consistency.
\end{enumerate}

\subsection{Storage Architecture}\label{s:hps-components}
Our HPS is implemented as a 3-level hierarchical memory architecture (\emph{cf.} \autoref{f:ec-async-update}) that utilizes GPU GDDR and/or high-bandwidth memory (HBM), distributed CPU memory and local SSD storage resources. The communication mechanisms between these components ensure that the most frequently used embeddings reside in the GPU embedding cache. Somewhat frequently used embeddings are cached in CPU memory, while a full copy of all model parameters, including those that rarely occur, is always kept available on the hard disk/SSD. To minimize delays, we overlap parameter updating and the migration of missing parameters from higher storage levels (SSD $\rightarrow$ CPU memory $\rightarrow$ GPU memory) with the dense model computation. The three memory architecture levels of the HPS are defined as follows:

\begin{figure*}[tb]
    \begin{minipage}{0.475\hsize}%
        \centering%
        \includegraphics[width=\hsize,trim=88 120 1353 258,clip]{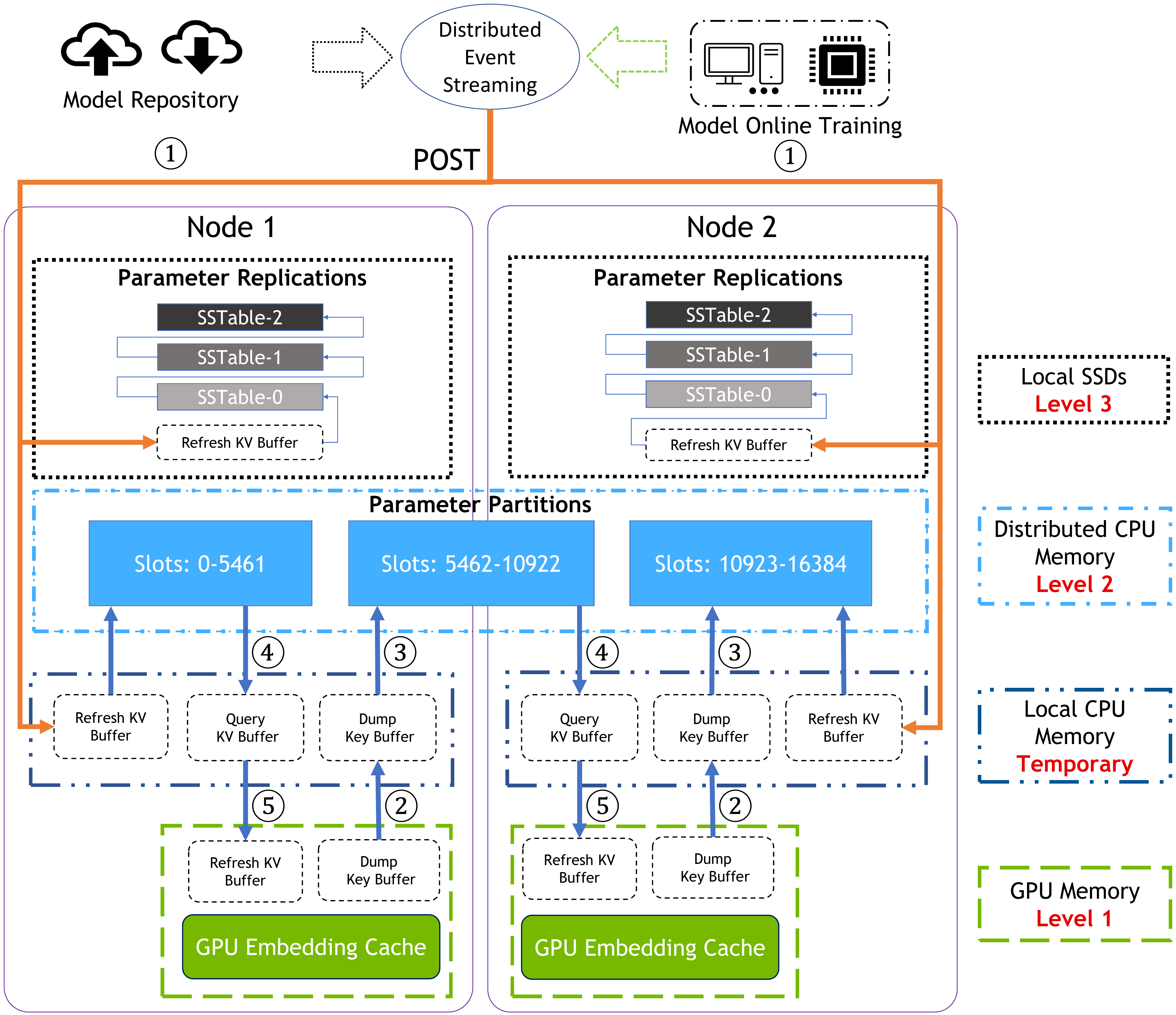}%
        \caption{HugeCTR HPS storage architecture. Annotated process (\textcircled{\raisebox{-0.1em}{1}}\textasciitilde\textcircled{\raisebox{-0.1em}{5}}) depicts asynchronous model updating and embedding cache refresh in a distributed deployment (see \autoref{s:update}).}%
        \label{f:ec-async-update}%
    \end{minipage}%
    \hfill%
    \begin{minipage}{0.475\hsize}%
        \pseudocode{a:ec-insert}{Embedding cache insertion.}{%
            \REQUIRE%
                $Q$ = query keys;
                $N$ = total size of embedding cache; 
                $t$ = hit rate threshold;
                $E_Q$ = pre-allocated buffer to store embeddings. 
            \STATE $ws$ $\gets$ \textsc{requestMemoryBlock}()  \COMMENT{Apply for workspace.}
            \STATE $Q^*$ $\gets$ \textsc{dedup}($Q$)  \COMMENT{Deduplicate keys.}
            \STATE \textsc{EmbCacheLookup}($Q^*$)  \COMMENT{GPU cache lookup keys}.
            \STATE $ws_\text{hitRate}$ $\gets$ $1 - \lvert ws_\text{missingKeys} \rvert \div N$
            \IF{$ws_\text{hitRate} < t$} 
                \STATE \textsc{ParServerLookup}($ws_\text{missingKeys}$)  \COMMENT{Synchronous!}
                \STATE \textsc{EmbCacheInsert}($ws_\text{missingKeys}$, $ws_\text{vectors}$) 
                \STATE $E_Q$ $\gets$ $ws_\text{vectors}$
                \STATE \textsc{free}($ws$)
            \ELSE
                \PROCEDURE{asyncInsert}{}
                    \STATE \textsc{ParServerLookup}($ws_\text{missingKeys}$)
                    \STATE \textsc{insert}($ws_\text{missingVectors}$)
                    \STATE \textsc{free}($ws$)
                \ENDPROCEDURE
                \STATE $E_Q$ $\gets$ $ws_\text{vectors}$   \COMMENT{Fill in default values.}
                \STATE Queue \textsc{asyncInsert}() task.
            \ENDIF
        }%
    \end{minipage}%
\end{figure*}

\textbf{GPU embedding cache} (level 1). This is a dynamic cache designed for recommendation model inference. It attempts to improve the lookup performance for embeddings by reducing additional/repetitive parameter movements through cleverly utilizing data locality to keep frequently used features (\emph{i.e.} the hot features) in the GPU memory. The GPU cache supports several operators (see \autoref{s:inf-ec}), as well as a dynamic insertion and an asynchronous refresh mechanism (see \autoref{s:update}) to retain a high cache hit rate.

\textbf{Parameter partitions} (level 2) store a partial copy of the embedding parameters in CPU memory. They act as an extension to the GPU embedding cache, and are queried if an embedding is required that is currently not present in the cache. Practitioners can choose between stand-alone deployment and cluster deployment, depending on their application scenario. In stand-alone deployment, the partitions are either placed in an optimized parallel hash-map (server-less deployment) or a local Redis instance. Distributed deployments can make use of multi-node Redis configurations. The contents of each partition are asynchronously adjusted in response to the queries processed by all inference nodes of a deployment. To receive online updates, parameter partitions can subscribe to topics from a distributed event stream.

\textbf{Parameter replications} (level 3). To ensure fault-tolerance, HPS retains a full copy of all model parameters (\emph{i.e.}, a model replica) in a disk-based RocksDB key-value store in each inference node. This fallback storage is accessed if a lookup request towards the corresponding \emph{parameter partitions} fails. Thus, if given enough time budget, a HPS deployment is always able to produce a full answer to every query. To stay up to date, each node separately monitors the \emph{distributed event stream} and applies online updates at its own pace.

\section{Inference GPU Embedding Cache}\label{s:inf-ec}

When processing online inference workloads, it is usually not possible to know which embedding table subsets will be required next. Therefore, our GPU embedding cache is designed as a general-purpose dynamic cache, which can accept new embeddings by evicting old embeddings. 

\subsection{Cache Data Model}\label{s:ec-structure}
\begin{figure*}[tb]
    \begin{minipage}[t]{0.475\hsize}%
        \centering%
        \includegraphics[width=0.85\hsize]{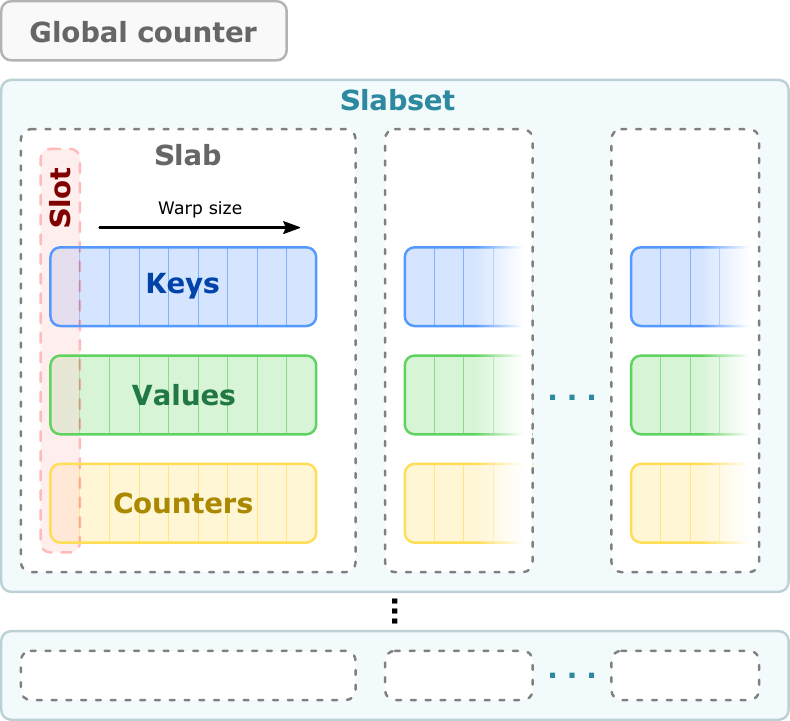}%
        \caption{GPU embedding cache data model.}%
        \label{f:ec-data-model}%
    \end{minipage}%
    \hfill%
    \begin{minipage}[t]{0.475\hsize}%
        \centering%
        \includegraphics[width=1\hsize]{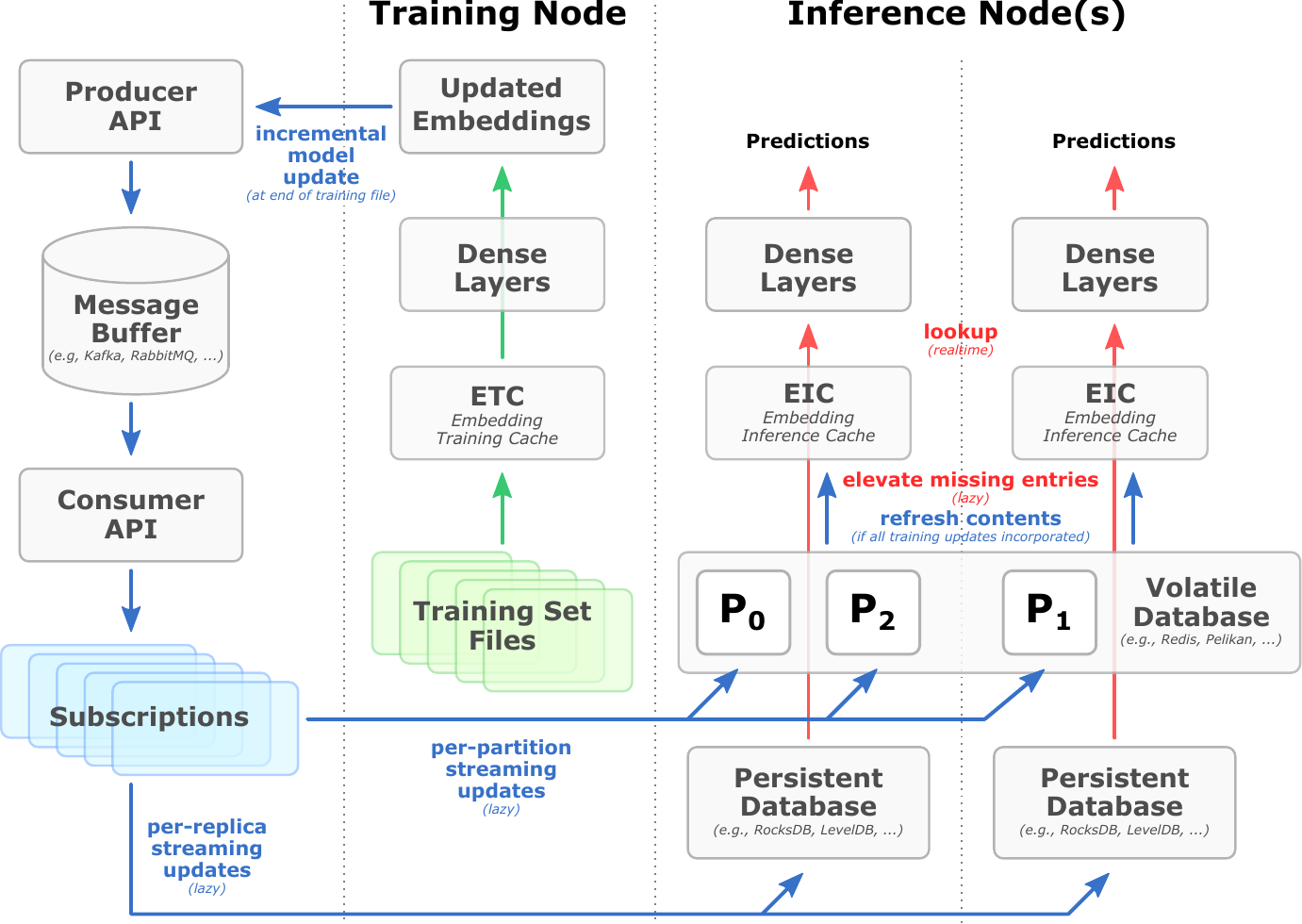}%
        \caption{General data-flow in HugeCTR model deployments.}%
        \label{f:general-data-flow}%
    \end{minipage}%
\end{figure*}

The GPU embedding cache consists of hierarchical 3-level structures as shown in \autoref{f:ec-data-model}: \emph{slots}, \emph{slabs} and \emph{slabsets}.

\textbf{Slots} represent the GPU embedding cache's basic storage unit. Each slot contains an embedding key, the associated embedding vector, and an access counter.

\textbf{Slabs.} Modern GPU architectures manage and execute code in warps (groups of 32 threads; \cite{CUDA_Warp}). Peak performance can be achieved by writing warp-aware programs. Therefore, we group 32 slots into one slab, so that each warp thread is assigned to a distinct slot. When searching for matching embedding keys, we use warps to linearly probe slabs. In determining if and where a key was found in a slab, we perform register-level intra-warp communications (shuffle, ballot, \emph{etc}.) to eliminate branch and memory divergences. 

\textbf{Slabsets.} Like cache lines are grouped into cache sets in N-way set-associative caches, slabs are packed into slabsets. To exploit the massively parallel computing power of GPUs, each embedding key is first mapped to a particular slabset, but may then occupy any slot in that slabset. This way, linear probing is confined to a single slabset, without conflicting with independent slabsets. A smaller slabset size can reduce key search latency, but also leads to increasing conflict misses. 
It is important to find the optimal slabset size to balance these two factors. We empirically set the slabset size to 2 for contemporary NVIDIA GPU architectures such as \emph{Ampere}. To maximize GPU resource utilization and inference concurrency, inference workers can share the same embedding cache. Race conditions are prevented by only granting a single warp exclusive access to a slabset for particular cache operations, such as query and replace. This approach also implicitly ensures thread safety. Because the total number of slabsets is usually much higher ($\ge$ millions) than the maximum number of warps per GPU ($\ge$ thousands), the mutual exclusion does not incur significant stalls.

\subsection{GPU Embedding Cache API}
The GPU embedding cache supports four APIs:
\begin{itemize}
    \item \textbf{Query} (\pseudoref{a:ec-query}) retrieves embedding vectors for sets of embedding keys. Missing keys are returned as a list that can be used to attempt fetching these embeddings from the parameter partitions.
    \item \textbf{Replace} (\pseudoref{a:ec-replace}) tries to insert embeddings by filling up empty slots first. If the number of empty slots is insufficient, the least recently used (LRU) embeddings are replaced. Already existing embeddings are ignored.
    \item \textbf{Update} (\pseudoref{a:ec-update}) first determines the intersection set between the input keys and the already cached keys, and then replaces the corresponding embedding vectors.
    \item \textbf{Dump} batch outputs the embedding keys currently stored in the cache.
\end{itemize}
\emph{Query}, \emph{Replace} and \emph{Update} share the same core algorithm (\emph{cf.} Algorithms \ref{a:ec-query}, \ref{a:ec-replace} and \ref{a:ec-update}). For each key, the assigned processing warp will first locate the slabset that contains the key using a hash function. Then it linearly probes the slabs within this slabset to either find the matching key-slot, or determine an empty/replaceable slot for insertion (replace \& update only). The \emph{Dump} API is trivial in that it simply copies all keys currently in the cache to the CPU memory.

\begin{figure*}[tb]
    \begin{minipage}{0.475\hsize}%
        \pseudocode{a:ec-query}{Embedding cache query.}{%
            \REQUIRE
            $Q$ = query keys (to lookup);
            $S$ = slabsets currently in GPU memory;
            $g$ = global iteration count.
            \ENSURE
            $R$ = retrieved embedding values;
            $M$ = indices in $Q$ and query key value of missed embeddings.
            \IF[Query API only]{global thread ID $=$ $0$}
                \STATE $g$ $\gets$ \textsc{atomicAdd}($g$, $1$)
            \ENDIF
            \STATE $workQueue$ $\gets$ \textsc{warpFetchKeys}($Q$, $tasksPerWarp$) 
            \WHILE{$workQueue$ $\neq$ $\emptyset$}
                \STATE $q$ $\gets$ \textsc{fetchNextKey}($workQueue$)  
                \STATE $s$ $\gets$ \textsc{slabsetHasher}($q$) $\mod$ $\lvert S \rvert$  \COMMENT{Find slabset.}
                \STATE $n_0$ $\gets$ \textsc{slabHasher}($q$) $\mod$ $\lvert s \rvert$  \COMMENT{Select first slab.}
                \STATE \textsc{lock}($s$)  \COMMENT{Ensure exclusive access to slabset.}
                \FOR[Iterate over slabset $s$.]{$i$ $\gets$ $0$ to $\lvert s \rvert$}
                    \hspace{-0.15em}\setlength{\fboxsep}{0.1em}\fcolorbox{RoyalBlue!50!White}{Cyan!5!White}{\parbox{\linewidth}{%
                    \vspace{0.15em}
                    \IF[Key missing.]{searched $\forall slab$ $\in$ $s$}
                        \STATE $M$ $\gets$ $M$ $\|$ $\langle \text{\textsc{indexOf}(}q\text{)}, q \rangle$  \COMMENT{Record it!}
                        \STATE \textbf{break for}
                    \ENDIF
                    \STATE $slab$ $\gets$ \textsc{warpReadOutSlab}($(n_0 + i) \mod \lvert s \rvert$)
                    \IF{$q$ $\in$ $slab$.\textsc{keys}}
                        \STATE \COMMENT{Key hit; get value, refresh counter!}
                        \STATE $R_{\text{\textsc{indexOf}(}q\text{)}}$ $\gets$ $slab$.\textsc{valueWhere}(\textsc{key} $=$ $q$)
                        \STATE $slab$.\textsc{counterWhere}(\textsc{key} $=$ $q$) $\gets$ $g$
                        \STATE \textbf{break for}
                    \ELSIF[Key missing.]{\textsc{Null} $\in$ $slab$.\textsc{keys}}
                        \STATE $M$ $\gets$ $M$ $\|$ $\langle \text{\textsc{indexOf}(}q\text{)}, q \rangle$  \COMMENT{Record it!}
                        \STATE \textbf{break for}
                    \ENDIF
                    \vspace{0.2em}
                    }}
                \ENDFOR
                \STATE \textsc{unlock}($s$)
            \ENDWHILE
        }
        \\[0.2em]
        \smaller Remark: $\|$ is a thread-safe concatenation operator.
    \end{minipage}%
    \hfill%
    \begin{minipage}{0.475\hsize}%
        \pseudocode{a:ec-replace}{Embedding cache replace.}{
            \REQUIRE
                $Q$ = keys (to be replaced);
                $E$ = associated embeddings.
            {\smaller\item[]\texttt{[$\cdots$ same as \pseudoref{a:ec-query} $\cdots$]}}
            \FOR[Iterate over slabset $s$.]{$i$ $\gets$ $0$ to $\lvert s \rvert$}
                \hspace{-0.15em}\setlength{\fboxsep}{0.1em}\fcolorbox{RoyalBlue!50!White}{Cyan!5!White}{\parbox{\linewidth}{%
                \vspace{0.15em}
                \IF[Key missing, slot full!]{searched $\forall slab$ $\in$ $s$}
                    \STATE $slot$ $\gets$ \textsc{warpFindSlotLRU}($s$, slot\_counter)
                    \STATE \textsc{warpWriteSlot}($slot$, $q$, $E_{q}$, $g$) \COMMENT{Replace!}
                    \STATE \textbf{break for}
                \ENDIF
                \STATE $slab$ $\gets$ \textsc{warpReadOutSlab}($(n_0 + i) \mod \lvert s \rvert$)
                \IF[Key hit; refresh counter!]{$q$ $\in$ $slab$.\textsc{keys}}
                    \STATE $slab$.\textsc{counterWhere}(\textsc{key} $=$ $q$) $\gets g$
                    \STATE \textbf{break for}
                \ELSIF[Slot available; insert!]{\textsc{Null} $\in$ $slab$.\textsc{keys}}
                    \STATE \textsc{warpWriteSlot}($emptySlot$, $q$, $E_{q}$, $g$)
                    \STATE \textbf{break for}
                \ENDIF
                \vspace{0.2em}
                }}%
            \ENDFOR
            {\smaller\item[]\texttt{[$\cdots$ same as \pseudoref{a:ec-query} $\cdots$]}}
        }%
        \\[0.8em]
        \pseudocode{a:ec-update}{Embedding cache update.}{%
            \REQUIRE
            $Q$ = keys (to be updated);
            $E$ = associated embeddings.
            {\smaller\item[]\texttt{[$\cdots$ same as \pseudoref{a:ec-query} $\cdots$]}}
            \FOR[Iterate over slabset $s$.]{$i$ $\gets$ $0$ to $\lvert s \rvert$}
                \hspace{-0.15em}\setlength{\fboxsep}{0.1em}\fcolorbox{RoyalBlue!50!White}{Cyan!5!White}{\parbox{\linewidth}{%
                \vspace{0.15em}
                \IF[Key not found; ignore!]{searched $\forall slab$ $\in$ $s$}
                    \STATE \textbf{break for}
                \ENDIF
                \STATE $slab$ $\gets$ \textsc{warpReadOutSlab}($(n_0 + i) \mod \lvert s \rvert$)
                \IF[Key hit; update value!]{$q$ $\in$ $slab$.\textsc{keys}}
                    \STATE $slab$.\textsc{valueWhere}(\textsc{key} $=$ $q$) $\gets$ $E_q$
                    \STATE \textbf{break for}
                \ELSIF[Not found, ignore!]{\textsc{Null} $\in$ $slab$.\textsc{keys}}
                    \STATE \textbf{break for}
                \ENDIF
                \vspace{0.2em}
                }}%
            \ENDFOR
            {\smaller\item[]\texttt{[$\cdots$ same as \pseudoref{a:ec-query} $\cdots$]}}
        }%
    \end{minipage}%
\end{figure*}

All APIs launch CUDA kernels that are executed asynchronously. \emph{I.e.}, the control flow is immediately returned to the CPU. Because they are thread safe on the slabset-level (see \autoref{s:ec-structure}), concurrent invocation of all APIs is permissible. To avoid frequent CUDA kernel launches and improve GPU resource utilization, all APIs accept mini-batches as input. The respective input keys are fairly distributed to warps, and pushed into a \emph{warp work queue}.

\subsection{Embedding Insertion}\label{s:ec-insert}
For failed lookups (\emph{i.e.}, key is currently not present in the GPU embedding cache), a cache insertion operation is triggered to fetch the missed embeddings from the parameter partitions in the CPU memory or a replica on a local SSD. As shown in \pseudoref{a:ec-insert}, the HPS has two insertion modes, between which the GPU embedding cache switches based on the relation between the current cache hit rate and a user-defined hit rate threshold:

\textbf{Asynchronous insertion} is activated if the cache hit rate is higher than the predefined threshold. For any missing keys, the default embedding vectors whose values are user configurable are returned immediately. The actual embeddings are fetched asynchronously from higher-level storage into the GPU embedding cache to have them available for future queries. This lazy insertion mechanism ensures that the prediction accuracy loss is negligible with a high hit rate.
    
\textbf{Synchronous insertion} blocks the rest of the pipeline until the missed embeddings have been fetched. With a reasonable threshold, synchronous insertion usually occurs only during the warm-up stage, or after model updates.

\section{CPU Memory and SSD Storage Layers}\label{s:hps-db}

To process models that scale beyond GPU memory capacity, in addition to the GPU embedding cache (\autoref{s:inf-ec}), the HPS incorporates two additional layers in its storage hierarchy. These layers are constructed based on either system memory, SSDs or network storage, and are highly modularized to support various backend implementations. 

\textbf{Volatile database (VDB)} layers (level 2 in \autoref{f:ec-async-update}) reside in volatile memory such as system memory, that requires traversal through a NVLink or the PCIe bus to access it from a GPU. In comparison to GPU memory, system memory can be extended at lower costs. To grow even further, VDBs can take advantage of multiple, low-latency system memories on a inference cluster. For example, using our \emph{RedisClusterBackend} VDB template implementation, users can use distributed Redis instances as a storage backend for embeddings. Thus, VDB implementations can, but do not have to be limited to machine boundaries. To distribute the workload, VDBs organize embedding table storage in partitions. Partitions are non-overlapping subsets of an embedding table that are stored in the same physical location. They are sparsely populated in response to the inference queries processed by all nodes that share VDB access. The maximum size (=\emph{overflow margin}) and amount of partitions per embedding table are configurable, and subject to a trade-off. More smaller partitions allow for smoother load balancing, but each partition adds a small processing overhead. 

VDBs are operated as an asynchronous cache. If a GPU embedding cache reports missing keys, the HPS queries the VDB next. Analogous to the embedding cache, each VDB entry contains a timestamp indicating when the entry was last accessed. For embedding vectors that were successfully retrieved, the VDB asynchronously updates this timestamp after returning the result. 
Missed embedding vectors are scheduled for insertion into the VDB to accelerate potential future queries. Thereby, the partition assignment of each embedding is fixed and determined by the XXH64-hash value \cite{collet2014xxhash} of the key. Insertions happen asynchronously to not stall pending lookup processes, and subsequently fill up the VDB partitions. Per-partition eviction policies determine what should be done if a partition exceeds its overflow margin. We implement multiple eviction policies. For example, the \emph{evict oldest} policy finds and prunes infrequently accessed keys.

\textbf{Persistent database (PDB)} layers (level 3 in \autoref{f:ec-async-update}) use hard-disks or SSDs to permanently store entire embedding tables (\emph{i.e.}, all model parameters). As such, the PDB is helpful to improve the prediction accuracy for datasets that exhibit an extreme long-tail distribution. PDB layers can serve as backup and ultimate ground truth for any number of models. To avoid key collisions, PDB implementations form separate key namespaces for each unique embedding table.

Our template implementation maps embedding tables to column groups in a RocksDB database, stored on a local SSD in each inference node. Hence, the entire model data is replicated in each inference node. This way, we achieve maximum fault tolerance, because node failures will not impair the ability of other inference nodes to fully answer each query. Continued operation is possible, even if a failure in a neighbor node brings down an attached Redis VDB. Without the VDB as an intermediate cache, it can of course take somewhat longer until embedding vectors for missed keys are asynchronously migrated into GPU embedding cache (see also \autoref{s:performance-evaluation}). However, assuming the GPU embedding cache can retain a high-enough hit rate, clients should only witness minor deviations in inference performance.

\section{Online model updating}\label{s:update}

Thus far, we have described how the HPS organizes resources to enable inference with pretrained models. In \autoref{f:general-data-flow}, we have highlighted this portion of the data-flow graph in red (\textcolor{red}{$\rightarrow$}). However, there exist many scenarios where recommendations depend on recent information (\emph{e.g.}, user interactions in social networks). After completing a training epoch incremental updates have to be propagated to all inference nodes for improved recommendations. Our HPS achieves this functionality using a dedicated online updating mechanism.

\textbf{Volatile \& persistent database update.} Model training is resource intensive, and therefore conducted by a set of nodes that is distinct form the inference cluster. Training sets for HugeCTR models are split into files that maximize the locality in the embedding cache. The model is trained by sequentially loading these files into the cache and processing the training episodes. Our online updating mechanism wraps around HugeCTR model training. It is designed as an auxiliary process (blue [\textcolor{blue}{$\rightarrow$}] data-flow graph in \autoref{f:general-data-flow}) that can be turned on and off at any point in time.

Once training progress has been made, the training nodes dump their updates to an Apache Kafka-based message buffer \cite{sax2018kafka}. This is done via our \emph{Message Producer API}, which handles serialization, batching, and the organization of updates into distinct message queues for each embedding table. Inference nodes that have loaded the affected model can use the corresponding \emph{Message Source API} to discover and subscribe to these message queues. As indicated in \autoref{f:general-data-flow}, separate subscriptions can be created for different VDB partitions. This allows nodes that share a VDB to also share the update workload among them. If a node becomes unresponsive, its current assignment is shifted to other nodes.

Applying online updates inevitably adds overhead. Therefore, we allow updates to be consumed lazily by each node using a background process. The execution of the update process is aligned with other I/O requests. To control and adjust the impact on online inference, users can limit the update ingestion speed and frequency.

Through message buffer subscriptions, updates are guaranteed to be in order and complete. Hence, upon fully processing all pending messages (sync), the individual database levels are guaranteed to be consistent (\emph{i.e.}, we guarantee final consistency). The lazy nature in which we apply updates implies that slight inconsistencies during model update periods have to be expected. However, in practice this does not matter because learning rates for model retraining are usually very small. As long as the optimization process is reasonably smooth, the prediction performance should not diminish significantly \cite{bennun2019dist-deep-learning,langer2020ddls}. Note that the same assumption also underpins the working principle of the GPU embedding cache's query API, which returns default embedding values for missed keys if the hit rate criteria is met (see \autoref{s:inf-ec}). However, since no downtime is required to ingest updates, it is possible to achieve continuous model improvement, which makes HPS particularly suitable for usage with highly active data sources.

\textbf{Asynchronous GPU embedding cache refresh}. The GPU embedding cache needs to be readily available when an inference request arrives. Ongoing streaming of small updates from message buffers to the GPU embedding cache would create spontaneous GPU load-spikes that are hard to predict and could diminish response times. Thus, instead of ingesting updates directly from Kafka, we allow the GPU embedding cache to regularly poll the VDB/PDB for updates and replace embeddings if necessary. The refresh cycle is configurable to best fit the training schedule. When using online training, the GPU embedding cache can be configured to periodically (minutes, hours, \emph{etc}.) refresh its contents. When using offline training, refreshes are triggered through signals sent by the Triton model management API \cite{triton_model_management}. \autoref{f:ec-async-update} illustrates the entire sequence until a model update becomes effective in the GPU embedding cache:
\begin{enumerate}[label=\raisebox{0.67pt}{\textcircled{\raisebox{-0.67pt}{\arabic*}}}]
    \item Monitor message stream. Dispatch and apply updates to CPU memory partitions (VDB) and the SSD (PDB).
    \item Dump GPU embedding cache keys in batches (size is configurable) and write them into the \emph{dump key buffer}.
    \item Lookup embedding keys, written to the \emph{dump key buffer}, from the CPU memory partitions and/or the SDD,
    \item and copy the corresponding embedding key-vectors to the \emph{queried key-vector buffer}.
    \item Download the \emph{queried key-vector buffer} into the GPU device and refresh the GPU embedding cache.
\end{enumerate}

\section{Performance Evaluation}\label{s:performance-evaluation}

In this section, we showcase the performance of our HPS from several aspects, including the end-to-end inference throughput and latency. Further, we provide empirical analyses of GPU embedding cache and different database backends, and investigate the impact of online updates on the HPS performance.

\subsection{Experiment Setup}\label{s:cluster-setup}
Unless specified otherwise, all experiments are carried out on a cluster consisting of NVIDIA DGX A100 \cite{nvidia2022data-center} nodes. Each node is equipped with two AMD EPYC 7742 CPUs, 2~TB of CPU memory, eight NVSwitch-interconnected NVIDIA A100 GPUs with 80~GB GPU memory each, and eight Mellanox CX6 InfiniBand adapters for inter-node communication.

To demonstrate the HPS’s capabilities, we use two publicly available and two synthetically generated datasets. For the public datasets, we trained a DLRM model to obtain an embedding table which is then used for inference.

\textbf{Criteo 1~TB} \cite{criteo2014dataset} is a large publicly available log of user click behavior in response to ads, containing 13 dense features and 26 sparse features. The final embedding table amounts to \textasciitilde90~GB, with an embedding vector size of 128. 

\textbf{MovieLens} \cite{movielens-25m} is a small publicly available dataset containing movie recommendations. It contains 3 sparse features, of which one is a multi-hot feature. The 
embedding table is only \textasciitilde20~MB large (embedding vector size = 128).

\textbf{Synthetic dataset A} mimics the properties of \emph{Criteo 1~TB} dataset. However, the final embedding table amounts to 650~GB. In lieu of generating a huge training dataset to obtain the embedding, we generate the embedding first by randomly creating embedding vectors of size 128. Then, we use the resulting key range to generate an inference request dataset by randomly drawing keys from a power-law distribution with $\alpha$ = 1.2 (see \autoref{s:power-law}). 
In the resulting inference requests about 95\% of the embedding table lookups reference 10\% of the embedding table. 

\textbf{Synthetic dataset B} is created in a similar way as \emph{Synthetic dataset A}, but contains 9 dense features and 130 sparse features. Further, we decreased the number of unique keys in each sparse feature so that the embedding table size becomes 81~GB large (\emph{i.e.}, close to the size of embedding table for the Criteo 1~TB dataset).

\subsection{Inference performance}\label{s:perf-study}

\subsubsection{Single-GPU single-instance deployment on Triton}
In this section, we evaluate the performance of HPS running on the top of a NVIDIA Triton Inference Server \cite{nvidia2022hugectr-backend}, in comparison with a PyTorch CPU implementation. To measure inference performance, we utilize Triton's performance analyzer \cite{triton_perf_analyzer}. For all datasets, the size of DLRM model's trained dense weights is at most 10~MB, which can be easily loaded into either CPU or GPU memory.

\begin{table}[t]
    \caption{HPS configuration for Criteo 1~TB dataset.}
    \label{t:hps-configs-criteo-1tb}%
    \centering%
    \smaller\smaller%
    \renewcommand{\arraystretch}{1.15}%
    \setlength{\tabcolsep}{0.75em}%
    \begin{tabular}{rcc}
		\toprule
		\textbf{Components} & \textbf{Parameter} & \textbf{Value}   \\
		\midrule & & \\[-1.75em]\midrule
		\multirow{3}{*}{GPU Emb. Cache}
		&  GPU cache \%        &  0.5  \\
		&  Hit rate threshold  &  0.8  \\
		&  \# instances / GPU  &    1  \\
		\midrule
		\multirow{3}{*}{Volatile Database} 
		&  Type                &  Hash map  \\
		&  Initial cache rate  &       1.0  \\
		&  \# partitions       &        16  \\
		\bottomrule
	\end{tabular}
\end{table}

\begin{table}[t]
    \centering%
    \caption{Volatile and persistent database random insertion.}%
    \label{t:database-backend-insert-perf}%
    \smaller\smaller%
    \renewcommand{\arraystretch}{1.15}%
    \setlength{\tabcolsep}{0.75em}%
    \begin{tabular}{rrrr}%
        \toprule
        {\hfil\textbf{Capacity}\hfil}
        & {\hfil\textbf{HashMap}\hfil}
        & {\hfil\textbf{Redis}\hfil}
        & {\hfil\textbf{RocksDB}\hfil}  \\[-0.15em]
        {\hfil\smaller(GB)\hfil}
        & {\hfil\smaller(VDB, 32 threads)\hfil}
        & {\hfil\smaller(VDB, 3 nodes)\hfil} 
        & {\hfil\smaller(PDB, SSD disk)\hfil}  \\
        \midrule & & & \\[-1.75em]\midrule
         10\hspace{1.5em}  &  245.3 MB/s\hspace{0.9em}  &  162.0 MB/s\hspace{0.4em}  &  52.1 MB/s\hspace{0.9em}  \\
         25\hspace{1.5em}  &  236.0 MB/s\hspace{0.9em}  &  138.9 MB/s\hspace{0.4em}  &  45.3 MB/s\hspace{0.9em}  \\
         50\hspace{1.5em}  &  237.4 MB/s\hspace{0.9em}  &  132.3 MB/s\hspace{0.4em}  &  40.2 MB/s\hspace{0.9em}  \\
         75\hspace{1.5em}  &  222.0 MB/s\hspace{0.9em}  &  120.9 MB/s\hspace{0.4em}  &  38.3 MB/s\hspace{0.9em}  \\
        100\hspace{1.5em}  &  203.9 MB/s\hspace{0.9em}  &  111.2 MB/s\hspace{0.4em}  &  37.4 MB/s\hspace{0.9em}  \\
        \bottomrule
    \end{tabular}%
\end{table}

\autoref{t:hps-configs-criteo-1tb} lists the configuration for the GPU embedding cache and the VDB. Frequent embeddings are kept in GPU memory. Since our test system has 2~TB CPU memory, we can increase the VDB capacity to fully cache the embedding table. In contrast, our PyTorch baseline keeps both the embedding table and the dense weights in the CPU memory.

\begin{table}[t]
    \centering%
    \caption{Embedding cache refreshment.}%
    \label{t:ec-refresh-table}%
    \smaller\smaller%
    \renewcommand{\arraystretch}{1.15}%
    \setlength{\tabcolsep}{0.75em}%
    \begin{tabular}{rrrr}%
        \toprule
        {\hfil\textbf{Capacity}\hfil}
        & {\hfil\textbf{Update}\hfil}
        & {\hfil\textbf{Dump}\hfil}
        & {\hfil\textbf{Bandwidth}\hfil}  \\[-0.15em]
        {\hfil\smaller(GB)\hfil}
        & {\hfil\smaller(ms)\hfil} 
        & {\hfil\smaller(ms)\hfil}
        & {\hfil\smaller(GB/s)\hfil}  \\
        \midrule & & & \\[-1.75em]\midrule
         1\hspace{1.5em}  &    5.152  &  0.064\hspace{0.2em}  &  194.20\hspace{1.2em}  \\
         5\hspace{1.5em}  &   25.177  &  0.184\hspace{0.2em}  &  198.62\hspace{1.2em}  \\
        10\hspace{1.5em}  &   50.262  &  0.335\hspace{0.2em}  &  198.96\hspace{1.2em}  \\
        20\hspace{1.5em}  &  100.530  &  0.641\hspace{0.2em}  &  198.95\hspace{1.2em}  \\
        40\hspace{1.5em}  &  200.345  &  1.19\hspace{0.2em}   &  199.73\hspace{1.2em}  \\
        \bottomrule
    \end{tabular}
\end{table}

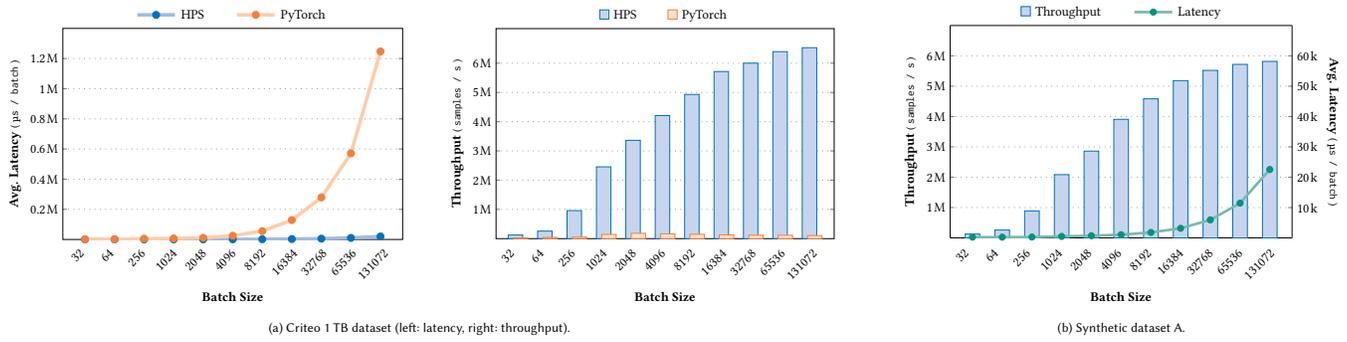
\begin{figure*}[tb]
    \newcommand{\figwidth}{7cm}%
    \newcommand{\figheight}{4.95cm}%
    \smaller\smaller\smaller%
    \begin{subfigure}[t]{0.62\hsize}%
        \centering%
        \resizebox{0.48\hsize}{!}{\begin{tikzpicture}%
            \begin{axis}[
                width = \figwidth, height = \figheight,
                xlabel = {\textbf{Batch Size}},
                xtick = data, symbolic x coords = {%
                    32, 64, 256, 1024, 2048, 4096, 8192, 16384, 32768, 65536, 131072%
                },
                xtick style = {draw=none},
                x tick label style={rotate=45,xshift=-0.5em,yshift=0.3em},
                enlarge x limits = 0.075,
                ylabel style = {align=center},
                ylabel = {\textbf{Avg. Latency} \smaller ( \texttt{{\textmu}s / batch} )},
                ymin = 0, ymax = 1400,
                ytick = {200, 400, 600, 800, 1000, 1200},
                yticklabels = {%
                    0.2\hspace{0.1em}M,
                    0.4\hspace{0.1em}M,
                    0.6\hspace{0.1em}M,
                    0.8\hspace{0.1em}M,
                    1\hspace{0.1em}M,
                    1.2\hspace{0.1em}M
                },
                ymajorgrids = true,
                major tick length=0.3em,
                major grid style={thin,dotted,Gray},
                legend style = {
                    at={(0.5,1.015)}, anchor=south,
                    draw=none,
                    inner sep=0,
                    legend columns=-1,
                    /tikz/every even column/.append style={column sep=2em}
                },
                legend cell align = {left}
            ]%
                \addlegendentry{HPS};
                \addplot[
                    color=LineColorA!20!ShadeColorA,ultra thick,
                    mark=*,mark options={LineColorA, scale=0.5}
                ] coordinates {
                    (32,      0.258)[ 0.258]
                    (64,      0.248)[ 0.248]
                    (256,     0.267)[ 0.267]
                    (1024,    0.417)[ 0.417]
                    (2048,    0.609)[ 0.609]
                    (4096,    0.973)[ 0.973]
                    (8192,    1.661)[ 1.661]
                    (16384,   2.867)[ 2.867]
                    (32768,   5.460)[ 5.460]
                    (65536,  10.248)[10.248]
                    (131072, 20.072)[20.072]
                };
                \addlegendentry{PyTorch};
                \addplot[
                    color=LineColorB!20!ShadeColorB,ultra thick,
                    mark=*,mark options={LineColorB,scale=0.5}
                ] coordinates {
                    (32,        1.313)[   1.313]
                    (64,        1.508)[   1.508]
                    (256,       4.975)[   4.975]
                    (1024,      7.130)[   7.130]
                    (2048,     11.185)[  11.185]
                    (4096,     24.852)[  24.852]
                    (8192,     55.278)[  55.278]
                    (16384,   128.639)[ 128.639]
                    (32768,   278.854)[ 278.854]
                    (65536,   571.534)[ 571.534]
                    (131072, 1246.993)[1246.993]
                };
            \end{axis}%
        \end{tikzpicture}}%
        \hfill%
        \resizebox{0.4675\hsize}{!}{\begin{tikzpicture}%
            \begin{axis}[
                width = \figwidth, height = \figheight,
                ybar = -0.7em,
                bar width = 1.1em,
                xlabel = {\textbf{Batch Size}},
                xtick = data, symbolic x coords = {%
                    32, 64, 256, 1024, 2048, 4096, 8192, 16384, 32768, 65536, 131072%
                },
                xtick style = {draw=none},
                x tick label style={rotate=45,xshift=-0.5em,yshift=0.6em},
                enlarge x limits = 0.075,
                ylabel style = {align=center},
                ylabel = {\textbf{Throughput} \smaller ( \texttt{samples / s} )},
                ymin = 0,
                ytick = {1000, 2000, 3000, 4000, 5000, 6000},
                yticklabels = {%
                    1\hspace{0.1em}M,
                    2\hspace{0.1em}M,
                    3\hspace{0.1em}M,
                    4\hspace{0.1em}M,
                    5\hspace{0.1em}M,
                    6\hspace{0.1em}M
                },
                ymajorgrids = true,
                major tick length=0.3em,
                major grid style={thin,dotted,Gray},
                point meta = explicit symbolic,
                legend style = {
                    at={(0.5,1.015)}, anchor=south,
                    draw=none,
                    inner sep=0,
                    legend columns=-1,
                    /tikz/every even column/.append style={column sep=2em}
                },
                legend image code/.code={%
                    \draw[#1,yshift=-0.3em] (0em,0em) rectangle (0.7em,0.7em);
                },
                legend cell align = {left}
            ]%
                \addlegendentry{HPS};
                \addplot[draw=LineColorA,fill=ShadeColorA,text=LineColorA] coordinates {
                    (32,      124.031)[ 124.031]
                    (64,      258.064)[ 258.064]
                    (256,     958.801)[ 958.801]
                    (1024,   2455.635)[2455.635]
                    (2048,   3362.889)[3362.889]
                    (4096,   4209.660)[4209.660]
                    (8192,   4931.968)[4931.968]
                    (16384,  5714.684)[5714.684]
                    (32768,  6001.465)[6001.465]
                    (65536,  6395.003)[6395.003]
                    (131072, 6530.091)[6530.091]
                };
                \addlegendentry{PyTorch};
                \addplot[draw=LineColorB,fill=ShadeColorB,text=LineColorB] coordinates {
                    (32,      24.354)[ 24.354]
                    (64,      42.423)[ 42.423]
                    (256,     51.448)[ 51.448]
                    (1024,   143.607)[143.607]
                    (2048,   183.097)[183.097]
                    (4096,   164.811)[164.811]
                    (8192,   148.194)[148.194]
                    (16384,  127.364)[127.364]
                    (32768,  117.509)[117.509]
                    (65536,  114.666)[114.666]
                    (131072, 105.110)[105.110]
                };
            \end{axis}%
        \end{tikzpicture}}%
        \caption{Criteo 1~TB dataset (left: latency, right: throughput).}%
        \label{f:perf_sg_si_criteo}%
    \end{subfigure}%
    \hfill%
    \begin{subfigure}[t]{0.331\hsize}%
        \centering%
        \resizebox{\hsize}{!}{\begin{tikzpicture}%
            \begin{axis}[
                width = \figwidth, height = \figheight,
                ybar = 0.45em,
                bar width = 1.1em,
                xlabel = {\textbf{Batch Size}},
                xtick = data, symbolic x coords = {%
                    32, 64, 256, 1024, 2048, 4096, 8192, 16384, 32768, 65536, 131072%
                },
                xtick style = {draw=none},
                x tick label style={rotate=45,xshift=-0.5em,yshift=0.6em},
                enlarge x limits = 0.075,
                axis y line* = left,
                ylabel = {\textbf{Throughput} \smaller ( \texttt{samples / s} )},
                ymin = 0, ymax = 7000,
                ytick = {1000, 2000, 3000, 4000, 5000, 6000},
                yticklabels = {%
                    1\hspace{0.1em}M,
                    2\hspace{0.1em}M,
                    3\hspace{0.1em}M,
                    4\hspace{0.1em}M,
                    5\hspace{0.1em}M,
                    6\hspace{0.1em}M
                },
                ymajorgrids = true,
                major tick length=0.3em,
                major grid style={thin,dotted,Gray},
                point meta = explicit symbolic,
                legend style = {
                    at={(0.33,1.015)}, anchor=south,
                    draw=none,
                    inner sep=0,
                    legend columns=-1,
                    /tikz/every even column/.append style={column sep=2em}
                },
                legend image code/.code={%
                    \draw[#1,yshift=-0.3em] (0em,0em) rectangle (0.7em,0.7em);
                }
            ]%
                \addlegendentry{Throughput};
                \addplot[draw=LineColorA,fill=ShadeColorA,text=LineColorA] coordinates {
                    (32,      128.000)[ 128.000]
                    (64,      259.109)[ 259.109]
                    (256,     888.888)[ 888.888]
                    (1024,   2085.539)[2085.539]
                    (2048,   2860.335)[2860.335]
                    (4096,   3908.396)[3908.396]
                    (8192,   4584.219)[4584.219]
                    (16384,  5183.169)[5183.169]
                    (32768,  5521.145)[5521.145]
                    (65536,  5719.172)[5719.172]
                    (131072, 5816.890)[5816.890]
                };
            \end{axis}
            \begin{axis}[
                width = \figwidth, height = \figheight,
                axis x line = none,
                xlabel = {\textbf{Batch size}},
                xtick = data, symbolic x coords = {%
                    32, 64, 256, 1024, 2048, 4096, 8192, 16384, 32768, 65536, 131072%
                },
                xtick style = {draw=none},
                enlarge x limits = 0.075,
                axis y line* = right,
                ylabel style = {align=center},
                ylabel = {\textbf{Avg. Latency} \smaller ( \texttt{{\textmu}s / batch} )},
                ylabel style = {align=center, rotate=-180},
                ymin = 0, ymax = 70,
                ytick = {10, 20, 30, 40, 50, 60},
                yticklabels = {%
                    10\hspace{0.1em}k,
                    20\hspace{0.1em}k,
                    30\hspace{0.1em}k,
                    40\hspace{0.1em}k,
                    50\hspace{0.1em}k,
                    60\hspace{0.1em}k
                },
                major tick length=0.3em,
                legend style = {
                    at={(0.67,1.015)}, anchor=south,
                    draw=none,
                    inner sep=0,
                    legend columns=-1,
                    /tikz/every even column/.append style={column sep=2em}
                }
            ]%
                \addlegendentry{Latency};
                \addplot[
                    color=LineColorC!50!White,very thick,
                    mark=*,mark options={LineColorC!90!White, scale=0.5}
                ] coordinates {
                    (32,      0.250)[ 0.250]
                    (64,      0.247)[ 0.247]
                    (256,     0.288)[ 0.288]
                    (1024,    0.491)[ 0.491]
                    (2048,    0.716)[ 0.716]
                    (4096,    1.048)[ 1.048]
                    (8192,    1.787)[ 1.787]
                    (16384,   3.161)[ 3.161]
                    (32768,   5.935)[ 5.935]
                    (65536,  11.459)[11.459]
                    (131072, 22.533)[22.533]
                };
            \end{axis}%
        \end{tikzpicture}}%
        \caption{Synthetic dataset A.}%
        \label{f:perf_sg_si_synth}%
    \end{subfigure}%
    \caption{Single-instance performance.}%
    \label{f:perf_sg}%
\end{figure*}

\autoref{f:perf_sg_si_criteo} compares the end-to-end inference performance of our HPS and PyTorch CPU \cite{torch-infer-cpu} with the \emph{Criteo 1~TB} dataset. We measure latency and throughput, while varying the batch size from 32 to 131,072. HPS significantly outperforms PyTorch CPU in terms of average latency per batch. Because GPU compute and memory resources can be exploited better, larger batch sizes lead to higher speedup. At the maximum batch size of 131,072, a 62x speedup is achieved. The throughput ranges from 2.4 million samples per second (batch size = 1,024) to 6.4 million samples per second (batch size = 131,072). In contrast, PyTorch CPU delivers at most 0.2 million samples per second (batch size = 2,048). It is also worth noting that HPS also has a 2.35x throughput advantage when compared to a TensorFlow GPU inference solution \cite{tf-infer-gpu} at batch size = 2,048 (1.43 million samples per second; model size = 15.6~GB).

For the \emph{Synthetic dataset A} we keep all settings the same, except the GPU cache percentage, which we lowered to 5\%, so that up to 32.5~GB of embeddings reside in GPU memory. Results are shown in \autoref{f:perf_sg_si_synth}. Like with Criteo, the throughput increases with the batch size and saturates around 131,072, while the latency remains stable.

\subsubsection{Multi-GPU multi-instance deployment on Triton}
To demonstrate how HPS can take advantage of multi-GPU environments, we set the batch size to 1,024 and test with both, \emph{Criteo 1~TB} and the \emph{synthetic dataset A}. We program the Triton Inference Server to evenly distribute inference instances while varying the number of GPUs \cite{triton_instance_group}. \autoref{f:perf_mg_mi} shows the resulting average QPS. With a single GPU, the QPS improves until up to 4 instances are deployed. This is due to enhanced GPU resource utilization from sharing the GPU embedding cache concurrently. Beyond 4 instances, increased resource contention degrades the QPS. Contention can be amortized by deploying the same number of instances on more GPUs. Consequently, the highest QPS (7.2x speedup) is achieved when deploying 8 model instances on 8 GPUs, so that each GPU has its own embedding cache. Note that they may still share VDB parameter partitions. To summarize, when deploying multiple instances, 
using both scaling up per GPU and multi-GPU scale out maximizes the QPS.

\subsubsection{Warm-up and stable stage performance of the GPU embedding cache}
To achieve stable performance, the HPS has to pass the warm-up stage, during which hot embeddings are fetched into the GPU embedding cache. During the warm-up stage, the hit rate keeps increasing until the cache is fully occupied. Thereby, the hit rate threshold controls whether cache updates should be applied either in synchronous or asynchronous mode (see \autoref{s:inf-ec}).

\begin{figure*}[tb]
    \newcommand{\figwidth}{7cm}%
    \newcommand{\figheight}{4.95cm}%
    \smaller\smaller\smaller%
    \begin{subfigure}[t]{0.32\hsize}%
        \centering%
        \resizebox{\hsize}{!}{\begin{tikzpicture}%
            \begin{axis}[
                width = \figwidth, height = \figheight,
                xlabel = {\textbf{Iteration \#}},
                enlarge x limits = 0.025,
                axis y line* = left,
                ylabel style = {align=center},
                ylabel = {\textbf{Latency} \smaller ( \texttt{ms} )},
                ymode = log,
                ymin = 0.4, ymax=20,
                ymajorgrids = true,
                major tick length=0.3em,
                major grid style={thin,dotted,Gray},
                legend style = {
                    at={(0.5,1.015)}, anchor=south,
                    draw=none,
                    inner sep=0,
                    legend columns=2,
                    /tikz/every even column/.append style={column sep=1.4em}
                },
                legend cell align = {left}
            ]%
                \addlegendentry{Latency (\texttt{threshold = 0})};
                \addplot[
                    color=LineColorA,very thick
                ] table [x=x, y=lat-0, col sep=comma] {results-7-2-3-crit.csv};
                \addlegendentry{Hit rate (\texttt{threshold = 0})};
                \addplot[color=LineColorD,very thick] coordinates { (0, 0) (0.0001, 0.0001) };
                \addlegendentry{Latency (\texttt{threshold = 0.5})};
                \addplot[
                    color=LineColorB,very thick
                ] table [x=x, y=lat-0-5, col sep=comma] {results-7-2-3-crit.csv};
                \addlegendentry{Hit rate (\texttt{threshold = 0.5})};
                \addplot[color=LineColorE,very thick] coordinates { (0, 0) (0.0001, 0.0001) };
                \addlegendentry{Latency (\texttt{threshold = 1})};
                \addplot[
                    color=LineColorC,very thick
                ] table [x=x, y=lat-1, col sep=comma] {results-7-2-3-crit.csv};
                \addlegendentry{Hit rate (\texttt{threshold = 1})};
                \addplot[color=LineColorF,very thick] coordinates { (0, 0) (0.0001, 0.0001) };
            \end{axis}%
            \begin{axis}[
                width = \figwidth, height = \figheight,
                axis x line = none,
                enlarge x limits = 0.025,
                axis y line* = right,
                ylabel style = {align=center, rotate=-180},
                ylabel = {\textbf{Hit Rate} \smaller ( \texttt{\%} )},
                ymin = 0, ymax = 0.9,
                ytick = {0.2, 0.4, 0.6, 0.8},
                yticklabels = {20, 40, 60, 80},
                ymajorgrids = true,
                major tick length=0.3em,
                major grid style={thin,loosely dashed,Gray}
            ]%
                \addplot[
                    color=LineColorD,very thick
                ] table [x=x, y=hit-0, col sep=comma] {results-7-2-3-crit.csv};
                \addplot[
                    color=LineColorE,very thick
                ] table [x=x, y=hit-0-5, col sep=comma] {results-7-2-3-crit.csv};
                \addplot[
                    color=LineColorF,very thick
                ] table [x=x, y=hit-1, col sep=comma] {results-7-2-3-crit.csv};
            \end{axis}
        \end{tikzpicture}}%
        \caption{Warm-up inference latency for the Criteo 1~TB dataset.}%
        \label{f:7.2.3.1}%
    \end{subfigure}%
    \hfill%
    \begin{subfigure}[t]{0.32\hsize}%
        \centering%
        \resizebox{\hsize}{!}{\begin{tikzpicture}%
            \begin{axis}[
                width = \figwidth, height = \figheight,
                xlabel = {\textbf{Iteration \#}},
                enlarge x limits = 0.025,
                axis y line* = left,
                ylabel style = {align=center},
                ylabel = {\textbf{Latency} \smaller ( \texttt{ms} )},
                ymode = log,
                ymin = 0.4, ymax=20,
                ymajorgrids = true,
                major tick length=0.3em,
                major grid style={thin,dotted,Gray},
                legend style = {
                    at={(0.5,1.015)}, anchor=south,
                    draw=none,
                    inner sep=0,
                    legend columns=2,
                    /tikz/every even column/.append style={column sep=1.4em}
                },
                legend cell align = {left}
            ]%
                \addlegendentry{Latency (\texttt{threshold = 0})};
                \addplot[
                    color=LineColorA,very thick
                ] table [x=x, y=lat-0, col sep=comma] {results-7-2-3-syn.csv};
                \addlegendentry{Hit rate (\texttt{threshold = 0})};
                \addplot[color=LineColorD,very thick] coordinates { (0, 0) (0.0001, 0.0001) };
                \addlegendentry{Latency (\texttt{threshold = 0.5})};
                \addplot[
                    color=LineColorB,very thick
                ] table [x=x, y=lat-0-5, col sep=comma] {results-7-2-3-syn.csv};
                \addlegendentry{Hit rate (\texttt{threshold = 0.5})};
                \addplot[color=LineColorE,very thick] coordinates { (0, 0) (0.0001, 0.0001) };
                \addlegendentry{Latency (\texttt{threshold = 1})};
                \addplot[
                    color=LineColorC,very thick
                ] table [x=x, y=lat-1, col sep=comma] {results-7-2-3-syn.csv};
                \addlegendentry{Hit rate (\texttt{threshold = 1})};
                \addplot[color=LineColorF,very thick] coordinates { (0, 0) (0.0001, 0.0001) };
            \end{axis}%
            \begin{axis}[
                width = \figwidth, height = \figheight,
                axis x line = none,
                enlarge x limits = 0.025,
                axis y line* = right,
                ylabel style = {align=center, rotate=-180},
                ylabel = {\textbf{Hit Rate} \smaller ( \texttt{\%} )},
                ymin = 0, ymax = 0.7,
                ytick = {0.2, 0.4, 0.6},
                yticklabels = {20, 40, 60},
                ymajorgrids = true,
                major tick length=0.3em,
                major grid style={thin,loosely dashed,Gray}
            ]%
                \addplot[
                    color=LineColorD,very thick
                ] table [x=x, y=hit-0, col sep=comma] {results-7-2-3-syn.csv};
                \addplot[
                    color=LineColorE,very thick
                ] table [x=x, y=hit-0-5, col sep=comma] {results-7-2-3-syn.csv};
                \addplot[
                    color=LineColorF,very thick
                ] table [x=x, y=hit-1, col sep=comma] {results-7-2-3-syn.csv};
            \end{axis}
        \end{tikzpicture}}%
        \caption{Warm-up inference latency for the Synthetic dataset A.}%
        \label{f:7.2.3.2}%
    \end{subfigure}%
    \hfill%
    \begin{subfigure}[t]{0.31\hsize}%
        \centering%
        \resizebox{\hsize}{!}{\begin{tikzpicture}%
            \begin{axis}[
                width = \figwidth, height = \figheight,
                axis y line* = right,
                xlabel = {\textbf{Interation \#}},
                xmin = 1025, xmax = 9993,
                xtick = {2048, 4096, 6144, 8192},
                x tick label style = {%
                    /pgf/number format/.cd,
                    scaled x ticks=false,
                    1000 sep={}
                },
                enlarge x limits = 0.025,
                ylabel = {\textbf{Hit Rate} \smaller ( \texttt{\%} )},
                ylabel style = {align=center, rotate=180},
                ymin = 0.3, ymax = 0.95,
                ytick = {0.4, 0.6, 0.8}, yticklabels = {40, 60, 80},
                major tick length=0.3em,
                ymajorgrids = true,
                major grid style={thin,dotted,Gray},
                legend style = {
                    at={(0.5,1.015)},
                    anchor=south,
                    draw=none,
                    inner sep=0,
                    legend columns=2,
                    /tikz/every even column/.append style={column sep=0.9em}
                },
                legend cell align = {left}
            ]
                \addlegendentry{Latency (\texttt{cache\% = 0.01})};
                \addplot[color=LineColorA,thick] coordinates { (0, 0) (1, 1) };
                \addlegendentry{Hit rate (\texttt{cache\% = 0.01})};
                \addplot[
                    color=LineColorD,thick
                ] table [x=idx, y=ds-hr-0.01, col sep=comma] {cache-rate-7-2.csv};
                \addlegendentry{Latency (\texttt{cache\% = 0.05})};
                \addplot[color=LineColorB,thick] coordinates { (0, 0) (1, 1) };
                \addlegendentry{Hit rate (\texttt{cache\% = 0.05})};
                \addplot[
                    color=LineColorE,thick
                ] table [x=idx, y=ds-hr-0.05, col sep=comma] {cache-rate-7-2.csv};
                \addlegendentry{Latency (\texttt{dlrm\_synthetic})};
                \addplot[color=LineColorC,thick] coordinates { (0, 0) (1, 1) };
                \addlegendentry{Hit rate (\texttt{dlrm\_synthetic})};
                \addplot[
                    color=LineColorF,thick
                ] table [x=idx, y=hr-syn, col sep=comma] {cache-rate-7-2.csv};
            \end{axis}
            \begin{axis}[
                width = \figwidth, height = \figheight,
                axis x line = none,
                axis y line* = left,
                xmin = 1025, xmax = 9993,
                enlarge x limits = 0.025,
                ylabel = {\textbf{Latency} \smaller ( \texttt{ms} )},
                ylabel style = {align=center},
                ymin = 0.5, ymax = 3.75,
                ytick = {1, 2, 3},
                major tick length=0.3em,
                ymajorgrids = true,
                major tick length=0.3em,
                major grid style={thin,loosely dashed,Gray}
            ]
                \addplot[
                    color=LineColorA,thick
                ] table [x=idx, y=ds-lat-0.01, col sep=comma] {cache-rate-7-2.csv};
                \addplot[
                    color=LineColorB,thick
                ] table [x=idx, y=ds-lat-0.05, col sep=comma] {cache-rate-7-2.csv};
                \addplot[
                    color=LineColorC,thick
                ] table [x=idx, y=lat-syn, col sep=comma] {cache-rate-7-2.csv};
            \end{axis}
        \end{tikzpicture}}%
        \caption{Stable stage inference latency for the Synthetic dataset A.}%
        \label{f:stable_stage_perf}%
    \end{subfigure}%
    \caption{Inference latency during warm-up and stable stage.}
    \label{f:inf-latency}
\end{figure*}
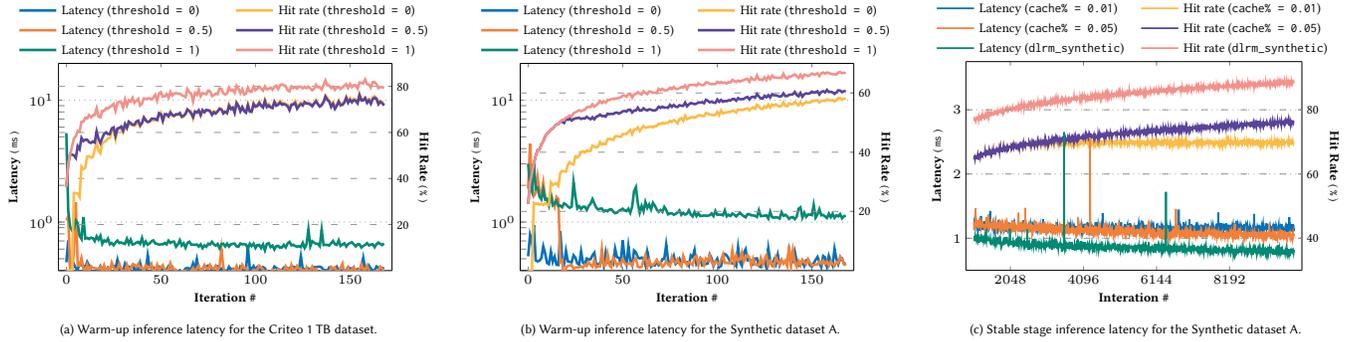

First, we study the GPU embedding cache's behavior during the warm-up stage. Figures \ref{f:7.2.3.1} (Criteo 1~TB) and \ref{f:7.2.3.2} (Synthetic dataset A) respectively show how the hit rate and inference latency change as the inference session progresses (batch size = 1,024). Using a hit rate threshold of 0.0, the inference latency stabilizes very quickly because the cache is always updated asynchronously (\emph{i.e.}, lazily). When setting the threshold to 1.0, the stabilization period is much longer. The overall latency is higher because cache updates block the inference pipeline. With a hit rate threshold of 0.5, the latency is at first relatively high because of blocking updates. Once the hit rate threshold is met, the latency is lowered and flattens overall. In other words, properly setting the hit rate threshold allows taking advantage of both blocking and asynchronous updates to balance the latency and hit rate.

Next, we analyze the further development of the latency as the cache enters the stable stage, where its hit rate saturates, using the synthetic inference request data. For this experiment, the batch size and hit rate threshold are both fixed at respectively 1,024 and 1.0. Results are shown in \autoref{f:stable_stage_perf}. We measured with GPU cache percentage ratios of 1\% and 5\%. Note that cutting the cache percentage by a fifth degrades the saturated hit rate from 76\% to 70\%, but leads only to a mean latency increase of 5\%. Thus, due to the HPS's effective usage of the request data's skewness property, a high performance can be retained, without having the embedding cache occupy too much GPU memory. The distribution of the input request data  also affects the cache performance. Using input request data with highly amplified locality that we generated just for this experiment (\texttt{dlrm\_synthetic} in \autoref{f:stable_stage_perf}) and a GPU cache percentage ratio of 5\%, the hit rate will eventually saturate at 100\%. The latency improves accordingly. As the hit rate surpasses 90\%, the latency is already 23\% lower than with the unaltered \emph{Synthetic dataset A}.

\subsubsection{End-to-end inference performance}
To verify its overall effectiveness, we study the end-to-end inference throughput performance of the HPS when combining differently configured storage layers (see \autoref{s:hps-db}). Results for the \emph{Criteo 1~TB} dataset are presented in \autoref{f:hps-e2e-infer-criteo}. We can draw the following conclusions: (1) HPS provides better inference performance with large batch sizes. (2) Unsurprisingly, the best performfance can be obtained by caching the entire embedding table. However, HPS still attains comparable performance figures for large models that cannot fit into memory. (3) Reducing the GPU cache ratio (20\% $\rightarrow$ 10\%), while increasing the maximum size of the Redis VDB (40\% $\rightarrow$ 45\%) yields better inference performance. Thus, the VDB as a 2nd-level cache can lower the pressure on the GPU cache, while the HPS-aware GPU cache update mechanism ensures a high hit rate, and greatly reduces the inference latency. 

The advantages of having a sophisticated hierarchical storage architecture become more remarkable as the model size increases. \autoref{f:hps-e2e-infer-synthetic} shows that granting the HPS just 5\% more CPU memory resources leads to an 1.24x end-to-end throughput increase with the much larger \emph{Synthetic dataset A}.

\begin{figure*}[tb]
    \centering
    \begin{minipage}[b]{0.475\hsize}%
        \newcommand{\figwidth}{7cm}%
        \newcommand{\figheight}{4.5cm}%
    
        \centering%
        \smaller\smaller\smaller%
        \resizebox{0.95\hsize}{!}{\begin{tikzpicture}%
            \begin{axis}[
                width = 10cm, height = 5cm,
                ybar = 0.3em,
                bar width = 0.8em,
                xlabel = {\textbf{Criteo 1~TB}\hspace{3.05cm}\textbf{Synthetic dataset A}},
                xtick = data, symbolic x coords = {%
                    crt-1, crt-2, crt-4, crt-8, syn-1, syn-2, syn-4, syn-8
                },
                xticklabels = {
                    1 GPU, 2 GPUs, 4 GPUs, 8 GPUs,
                    1 GPU, 2 GPUs, 4 GPUs, 8 GPUs
                },
                major tick length = 0,
                minor x tick num = 1,
                minor tick length = 1em,
                xtick pos = left,
                enlarge x limits = 0.1,
                ylabel style = {align=center},
                ylabel = {\textbf{QPS} \smaller (\texttt{batches / s})},
                ymin = 0, ymax = 6250,
                ytick = {1000, 3000, 5000},
                yticklabels = {1 k, 3 k, 5 k},
                ymajorgrids = true,
                major grid style={thin,dotted,Gray},
                point meta = explicit symbolic,
                nodes near coords = {
                    \pgfkeys{/pgf/fpu=true}%
                    \pgfmathparse{\pgfplotspointmeta>0}%
                    \ifpgfmathfloatcomparison%
                        \pgfmathprintnumber[1000 sep={},precision=0]{\pgfplotspointmeta}%
                    \fi%
                },
                nodes near coords align = {vertical},
                nodes near coords style={font=\tiny},
                legend style = {
                    at={(0.5,1.015)}, anchor=south,
                    draw=none,
                    inner sep=0,
                    legend columns=4,
                    /tikz/every even column/.append style={column sep=2em}
                },
                legend image code/.code={%
                    \draw[#1,yshift=-0.3em] (0em,0em) rectangle (0.7em,0.7em);
                },
                legend cell align={left}
            ]
                \addlegendentry{1 Instance};
                \addplot[draw=LineColorA, fill=ShadeColorA, text=LineColorA] coordinates {
                    (crt-1, 1700.4)[1700.4] 
                    (crt-2, -0.1)[0.0]
                    (crt-4, -0.1)[0.0]
                    (crt-8, -0.1)[0.0]
                    (syn-1, 1762.0)[1762.0] 
                    (syn-2, -0.1)[0.0]
                    (syn-4, -0.1)[0.0]
                    (syn-8, -0.1)[0.0]
                };
                \addlegendentry{2 Instances};
                \addplot[draw=LineColorB, fill=ShadeColorB, text=LineColorB] coordinates {
                    (crt-1, 2516.4)[2516.4] 
                    (crt-2, 2941.2)[2941.2] 
                    (crt-4, -0.1)[0.0]
                    (crt-8, -0.1)[0.0]
                    (syn-1, 2280.2)[2280.2] 
                    (syn-2, 2378.0)[2378.0] 
                    (syn-4, -0.1)[0.0]
                    (syn-8, -0.1)[0.0]
                };
                \addlegendentry{4 Instances};
                \addplot[draw=LineColorC, fill=ShadeColorC, text=LineColorC] coordinates {
                    (crt-1, 3235.4)[3235.4] 
                    (crt-2, 4046.4)[4046.4] 
                    (crt-4, 4317.2)[4317.2] 
                    (crt-8, -0.1)[0.0]
                    (syn-1, 3129.6)[3129.6] 
                    (syn-2, 4004.0)[4004.0] 
                    (syn-4, 4345.6)[4345.6] 
                    (syn-8, -0.1)[0.0]
                };
                \addlegendentry{8 Instances}
                \addplot[draw=LineColorD, fill=ShadeColorD, text=LineColorD] coordinates {
                    (crt-1,  787.4)[ 787.4] 
                    (crt-2, 1093.4)[1093.4] 
                    (crt-4, 4747.8)[4747.8] 
                    (crt-8, 5669.8)[5177.8] 
                    (syn-1, 2819.2)[2819.2] 
                    (syn-2, 2839.4)[2839.4] 
                    (syn-4, 5177.8)[5177.8] 
                    (syn-8, 5489.2)[5489.2] 
                };
            \end{axis}
            \draw(rel axis cs: 0.5835, 0.0) -- (rel axis cs: 0.5835, -0.125);
        \end{tikzpicture}}%
        \caption{Multi-GPU, multi-instance batch throughput.}%
        \label{f:perf_mg_mi}%
    \end{minipage}%
    \hfill%
    \begin{minipage}[b]{0.480\hsize}%
        \newcommand{\figwidth}{7cm}%
        \newcommand{\figheight}{4.25cm}%
        
        \centering%
        \smaller\smaller\smaller%
        \resizebox{0.8\hsize}{!}{\begin{tikzpicture}%
            \begin{axis}[
                width = \figwidth, height = \figheight,
                xlabel = {\textbf{Iteration \#}},
                enlarge x limits = 0.025,
                axis y line* = left,
                ylabel style = {align=center},
                ylabel = {\textbf{Accuracy} \smaller ( \texttt{\%} )},
                ymin = 0.2, ymax = 1,
                ytick = {0.4, 0.6, 0.8},
                yticklabels = {40, 60, 80},
                ymajorgrids = true,
                major tick length=0.3em,
                major grid style={thin,dotted,Gray},
                legend style = {
                    at={(0.5,1.015)}, anchor=south,
                    draw=none,
                    inner sep=0,
                    legend columns=2,
                    /tikz/every even column/.append style={column sep=1.4em}
                },
                legend cell align = {left}
            ]%
                \addlegendentry{Accuracy (\texttt{threshold = 0})};
                \addplot[
                    color=LineColorA,thick
                ] table [x=x, y=acc-0, col sep=comma] {acc-hit-rate-crit.csv};
                \addlegendentry{Hit rate (\texttt{threshold = 0})};
                \addplot[color=LineColorD,very thick] coordinates { (0, 0) (0.0001, 0.0001) };
                \addlegendentry{Accuracy (\texttt{threshold = 0.5})};
                \addplot[
                    color=LineColorB,thick,dashed
                ] table [x=x, y=acc-0-5, col sep=comma] {acc-hit-rate-crit.csv};
                \addlegendentry{Hit rate (\texttt{threshold = 0.5})};
                \addplot[color=LineColorE,very thick] coordinates { (0, 0) (0.0001, 0.0001) };
                \addlegendentry{Accuracy (\texttt{threshold = 1})};
                \addplot[
                    color=LineColorC,thick,dotted
                ] table [x=x, y=acc-1, col sep=comma] {acc-hit-rate-crit.csv};
                \addlegendentry{Hit rate (\texttt{threshold = 1})};
                \addplot[color=LineColorF,very thick] coordinates { (0, 0) (0.0001, 0.0001) };
            \end{axis}%
            \begin{axis}[
                width = \figwidth, height = \figheight,
                axis x line = none,
                enlarge x limits = 0.025,
                axis y line* = right,
                ylabel style = {align=center, rotate=-180},
                ylabel = {\textbf{Hit Rate} \smaller ( \texttt{\%} )},
                ymin = 0.2, ymax = 1,
                ytick = {0.4, 0.6, 0.8},
                yticklabels = {40, 60, 80},
                ymajorgrids = false,
                major tick length=0.3em,
                major grid style={thin,loosely dashed,Gray}
            ]%
                \addplot[
                    color=LineColorD,thick
                ] table [x=x, y=hit-0, col sep=comma] {acc-hit-rate-crit.csv};
                \addplot[
                    color=LineColorE,thick
                ] table [x=x, y=hit-0-5, col sep=comma] {acc-hit-rate-crit.csv};
                \addplot[
                    color=LineColorF,thick
                ] table [x=x, y=hit-1, col sep=comma] {acc-hit-rate-crit.csv};
            \end{axis}
        \end{tikzpicture}}%
        \caption{Criteo 1~TB: Accuracy and hit rate during inference.}%
        \label{f:acc_hit_rate}%
    \end{minipage}
\end{figure*}
    
\begin{figure*}[tb]
    \newcommand{\figwidth}{7cm}%
    \newcommand{\figheight}{5cm}%
    
    \centering
    \begin{subfigure}[t]{0.58\hsize}%
        \centering%
        \smaller\smaller\smaller%
        \resizebox{\hsize}{!}{\begin{tikzpicture}%
            \begin{axis}[
                width = 13cm, height = \figheight,
                ybar = 0.45em,
                bar width = 1.85em,
                xlabel = {\textbf{Batch Size}},
                xtick = data, symbolic x coords = {1024, 2048, 4096, 8192},
                xtick style = {draw=none},
                enlarge x limits = 0.18,
                ylabel style = {align=center},
                ylabel = {\textbf{Throughput} \smaller ( \texttt{samples / s} )},
                ymode = log, log basis y = 10,
                ymax = 4500,
                ytick = {250, 500, 1000, 2000},
                yticklabels = {250\hspace{0.1em}k, 500\hspace{0.1em}k, 1\hspace{0.1em}M, 2\hspace{0.1em}M},
                ymajorgrids = true,
                major tick length=0.3em,
                major grid style={thin,dotted,Gray},
                point meta = explicit symbolic,
                nodes near coords = {
                    \pgfkeys{/pgf/fpu=true}%
                    \pgfmathparse{\pgfplotspointmeta<1000}%
                    \ifpgfmathfloatcomparison%
                        \pgfmathprintnumber[precision=0]{\pgfplotspointmeta}\hspace{0.2em}k%
                    \else%
                        \pgfmathparse{\pgfplotspointmeta / 1000}%
                        \pgfmathprintnumber[precision=2]{\pgfmathresult}\hspace{0.2em}M%
                    \fi%
                },
                nodes near coords align = {vertical},
                nodes near coords style={font=\tiny},
                legend style = {
                    at={(1,1.015)}, anchor=south east,
                    draw=none,  
                    inner sep=0,
                    legend columns=-1,
                    /tikz/every even column/.append style={column sep=1em}
                },
                legend image code/.code={%
                    \draw[#1,yshift=-0.3em] (0em,0em) rectangle (0.7em,0.7em);
                }
            ]
                \addlegendentry{10\% EC + 45\% Redis + RocksDB};
                \addplot[draw=LineColorA, fill=ShadeColorA, text=LineColorA] coordinates {
                    (1024,  775.133711)[ 775.133711]
                    (2048,  869.640838)[ 869.640838]
                    (4096, 1126.760563)[1126.760563]
                    (8192, 1273.398701)[1273.398701]
                };
                \addlegendentry{20\% EC + 40\% Redis + RocksDB};
                \addplot[draw=LineColorB, fill=ShadeColorB, text=LineColorB] coordinates {
                    (1024, 536.020583)[ 536.020583]
                    (2048, 746.491490)[ 746.491490]
                    (4096, 753.012048)[ 753.012048]
                    (8192, 876.501008)[ 876.501008]
                };
                \addlegendentry{20\% EC + PHM};
                \addplot[draw=LineColorC, fill=ShadeColorC, text=LineColorC] coordinates {
                    (1024, 1166.316772)[1166.316772]
                    (2048, 1970.831691)[1970.831691]
                    (4096, 2766.251729)[2766.251729]
                    (8192, 3137.293623)[3137.293623]
                };
                \addlegendentry{20\% EC + Redis};
                \addplot[draw=LineColorD, fill=ShadeColorD, text=LineColorD] coordinates {
                    (1024,  838.293235)[ 838.293235]
                    (2048, 1194.600406)[1194.600406]
                    (4096, 1297.521733)[1297.521733]
                    (8192, 1701.170422)[1701.170422]
                };
                \addlegendentry{20\% EC + RocksDB};
                \addplot[draw=LineColorE, fill=ShadeColorE, text=LineColorE] coordinates {
                    (1024, 222.464461)[ 222.464461]
                    (2048, 246.445030)[ 246.445030]
                    (4096, 201.641361)[ 201.641361]
                    (8192, 259.652585)[ 259.652585]
                };
            \end{axis}
        \end{tikzpicture}}%
        \caption{Criteo 1~TB dataset; PHM = Parallel HashMap.}%
        \label{f:hps-e2e-infer-criteo}%
    \end{subfigure}%
    \hspace{0.05\hsize}%
    \begin{subfigure}[t]{0.30\hsize}%
        \centering%
        \smaller\smaller\smaller%
        \resizebox{\hsize}{!}{\begin{tikzpicture}%
            \begin{axis}[
                width = \figwidth, height = \figheight,
                ybar = 0.45em,
                bar width = 1.85em,
                xlabel = {\textbf{Batch Size}},
                xtick = data, symbolic x coords = {1024, 2048, 4096, 8192},
                xtick style = {draw=none},
                enlarge x limits = 0.18,
                ylabel style = {align=center},
                ylabel = {\textbf{Throughput} \smaller ( \texttt{samples / s} )},
                ymin = 350, ymax = 650,
                ytick = {400, 500, 600},
                yticklabels = {0.4\hspace{0.1em}M, 0.5\hspace{0.1em}M, 0.6\hspace{0.1em}M},
                ymajorgrids = true,
                major tick length=0.3em,
                major grid style={thin,dotted,Gray},
                point meta = explicit symbolic,
                nodes near coords = {\pgfmathprintnumber[precision=0]{\pgfplotspointmeta}\hspace{0.2em}k},
                nodes near coords align = {vertical},
                nodes near coords style={font=\tiny},
                legend style = {
                    at={(1,1.015)}, anchor=south east,
                    draw=none,
                    inner sep=0,
                    legend columns=-1,
                    /tikz/every even column/.append style={column sep=1em}
                },
                legend image code/.code={%
                    \draw[#1,yshift=-0.3em] (0em,0em) rectangle (0.7em,0.7em);
                }
            ]
                \addlegendentry{5\% EC + 20\% Redis + RocksDB};
                \addplot[draw=LineColorA, fill=ShadeColorA, text=LineColorA] coordinates {
                    (1024, 408.945687)[408.945687]
                    (2048, 458.987)[458.987]
                    (4096, 468.1142857)[468.114285]
                    (8192, 477.2780238)[477.278023]
                };
                \addlegendentry{5\% EC + 25\% Redis + RocksDB};
                \addplot[draw=LineColorB, fill=ShadeColorB, text=LineColorB] coordinates {
                    (1024, 502.206964)[502.206964]
                    (2048, 573.733751)[573.733751]
                    (4096, 584.308131)[584.308131]
                    (8192, 591.480144)[591.480144]
                };
            \end{axis}
        \end{tikzpicture}}%
        \caption{Synthetic dataset A.}%
        \label{f:hps-e2e-infer-synthetic}%
    \end{subfigure}%
    
    \caption{HPS end-to-end inference throughput. Redis refers to a VDB with 40 storage partitions spread across a 3-node Redis cluster.}%
    \label{f:hps-e2e}%
\end{figure*}
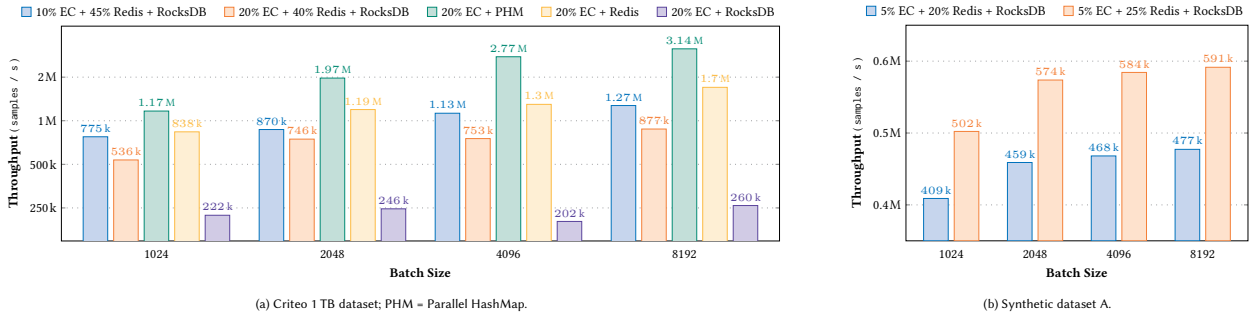

\begin{figure*}[tb]
    \newcommand{\figwidth}{7cm}%
    \newcommand{\figheight}{5cm}%
    
    \centering
    \begin{subfigure}[t]{0.32\hsize}%
        \centering%
        \smaller\smaller\smaller%
        \resizebox{\hsize}{!}{\begin{tikzpicture}%
            \begin{axis}[
                width = \figwidth, height = \figheight,
                ybar = 0.45em,
                bar width = 1.1em,
                xlabel = {\textbf{Batch Size}},
                xtick = data, symbolic x coords = {%
                    32, 64, 128, 256, 512, 1024, 2048, 4096, 8192, 16384, 32768, 65536%
                },
                xtick style = {draw=none},
                x tick label style={rotate=45,xshift=-0.5em,yshift=0.6em},
                enlarge x limits = 0.075,
                axis y line* = right,
                ylabel style = {align=center, rotate=-180},
                ylabel = {\textbf{Hit rate} \smaller ( \texttt{\%} )},
                ymin = 0.9825, ymax = 0.99625,
                ytick = {0.985,0.9875,0.990,0.9925,0.995},
                yticklabels = {%
                    98.5,%
                    98.75,%
                    99,%
                    99.25,%
                    99.5%
                },
                ymajorgrids = true,
                major tick length=0.3em,
                major grid style={thin,dotted,Gray},
                point meta = explicit symbolic,
                legend style = {
                    at={(0.11,1.015)}, anchor=south,
                    draw=none,
                    inner sep=0,
                    legend columns=-1,
                    /tikz/every even column/.append style={column sep=2em}
                },
                legend image code/.code={%
                    \draw[#1,yshift=-0.3em] (0em,0em) rectangle (0.7em,0.7em);
                }
            ]%
                \addlegendentry{Hit rate};
                \addplot[draw=LineColorD,fill=ShadeColorD,text=LineColorD] coordinates {
                    (32,    0.985042)[0.985042]
                    (64,    0.984914)[0.984914]
                    (128,   0.986809)[0.986809]
                    (256,   0.985998)[0.985998]
                    (512,   0.984488)[0.984488]
                    (1024,  0.984717)[0.984717]
                    (2048,  0.985610)[0.985610]
                    (4096,  0.985921)[0.985921]
                    (8192,  0.989462)[0.989462]
                    (16384, 0.990117)[0.990117]
                    (32768, 0.991396)[0.991396]
                    (65536, 0.990771)[0.990771]
                };
            \end{axis}
            \begin{axis}[
                width = \figwidth, height = \figheight,
                axis x line = none,
                xlabel = {\textbf{Batch size}},
                xtick = data, symbolic x coords = {%
                    32, 64, 128, 256, 512, 1024, 2048, 4096, 8192, 16384, 32768, 65536%
                },
                xtick style = {draw=none},
                enlarge x limits = 0.075,
                axis y line* = left,
                ylabel style = {align=center},
                ylabel = {\textbf{Avg. Latency} \smaller ( \texttt{ms / batch} )},
                ymin = 0, ymax = 5.5,
                ytick = {1, 2, 3, 4, 5},
                major tick length=0.3em,
                legend style = {
                    at={(0.62,1.015)}, anchor=south,
                    draw=none,
                    inner sep=0,
                    legend columns=-1,
                    /tikz/every even column/.append style={column sep=2em}
                }
            ]%
                \addlegendentry{T4};
                \addplot[
                    color=LineColorA!50!White,very thick,
                    mark=*,mark options={LineColorA!90!White, scale=0.5}
                ] coordinates {
                    (32,    0.19164)[0.19164]
                    (64,    0.241  )[0.241  ]
                    (128,   0.423  )[0.423  ]
                    (256,   0.437  )[0.437  ]
                    (512,   0.473  )[0.473  ]
                    (1024,  0.472  )[0.472  ]
                    (2048,  0.458  )[0.458  ]
                    (4096,  0.5825 )[0.5825 ]
                    (8192,  0.8463 )[0.8463 ]
                    (16384, 1.541  )[1.541  ]
                    (32768, 2.811  )[2.811  ]
                    (65536, 5.058  )[5.058  ]
                };
                \addlegendentry{A30};
                \addplot[
                    color=LineColorF!70!White,very thick,
                    mark=*,mark options={LineColorF, scale=0.5}
                ] coordinates {
                    (32,    0.168   )[0.168   ]
                    (64,    0.180826)[0.180826]
                    (128,   0.19384 )[0.19384 ]
                    (256,   0.211956)[0.211956]
                    (512,   0.231457)[0.231457]
                    (1024,  0.258   )[0.258   ]
                    (2048,  0.315314)[0.315314]
                    (4096,  0.449116)[0.449116]
                    (8192,  0.681746)[0.681746]
                    (16384, 1.241128)[1.241128]
                    (32768, 2.195967)[2.195967]
                    (65536, 3.903513)[3.903513]
                };
                \addlegendentry{A100};
                \addplot[
                    color=LineColorC!50!White,very thick,
                    mark=*,mark options={LineColorC!90!White, scale=0.5}
                ] coordinates {
                    (32,    0.167002)[0.167002]
                    (64,    0.183343)[0.183343]
                    (128,   0.20196 )[0.20196 ]
                    (256,   0.216294)[0.216294]
                    (512,   0.234857)[0.234857]
                    (1024,  0.281811)[0.281811]
                    (2048,  0.450771)[0.450771]
                    (4096,  0.541947)[0.541947]
                    (8192,  0.747654)[0.747654]
                    (16384, 1.279079)[1.279079]
                    (32768, 2.347651)[2.347651]
                    (65536, 3.633   )[3.633   ]
                };
            \end{axis}%
        \end{tikzpicture}}%
        \caption{MovieLens dataset}%
        \label{f:movielens}%
    \end{subfigure}%
    \hfill%
    \begin{subfigure}[t]{0.32\hsize}%
        \centering%
        \smaller\smaller\smaller%
        \resizebox{\hsize}{!}{\begin{tikzpicture}%
            \begin{axis}[
                width = \figwidth, height = \figheight,
                ybar = 0.45em,
                bar width = 1.1em,
                xlabel = {\textbf{Batch Size}},
                xtick = data, symbolic x coords = {%
                    32, 64, 128, 256, 512, 1024, 2048, 4096, 8192, 16384, 32768, 65536%
                },
                xtick style = {draw=none},
                x tick label style={rotate=45,xshift=-0.5em,yshift=0.6em},
                enlarge x limits = 0.075,
                axis y line* = right,
                ylabel style = {align=center, rotate=-180},
                ylabel = {\textbf{Hit rate} \smaller ( \texttt{\%} )},
                ymin = 0.905, ymax = 0.995,
                ytick = {0.92,0.935,0.95,0.965,0.98},
                yticklabels = {%
                    92,
                    93.5,%
                    95,%
                    96.5,%
                    98,%
                },
                ymajorgrids = true,
                major tick length=0.3em,
                major grid style={thin,dotted,Gray},
                point meta = explicit symbolic,
                legend style = {
                    at={(0.11,1.015)}, anchor=south,
                    draw=none,
                    inner sep=0,
                    legend columns=-1,
                    /tikz/every even column/.append style={column sep=2em}
                },
                legend image code/.code={%
                    \draw[#1,yshift=-0.3em] (0em,0em) rectangle (0.7em,0.7em);
                }
            ]%
                \addlegendentry{Hit rate};
                \addplot[draw=LineColorD,fill=ShadeColorD,text=LineColorD] coordinates {
                    (32,    0.9809667892)[0.9809667892]
                    (64,    0.9799750228)[0.9799750228]
                    (128,   0.9772611092)[0.9772611092]
                    (256,   0.9740196099)[0.9740196099]
                    (512,   0.9699425205)[0.9699425205]
                    (1024,  0.9643837864)[0.9643837864]
                    (2048,  0.9569936306)[0.9569936306]
                    (4096,  0.9470993394)[0.9470993394]
                    (8192,  0.9349061848)[0.9349061848]
                };
            \end{axis}
            \begin{axis}[
                width = \figwidth, height = \figheight,
                axis x line = none,
                xlabel = {\textbf{Batch size}},
                xtick = data, symbolic x coords = {%
                    32, 64, 128, 256, 512, 1024, 2048, 4096, 8192, 16384, 32768, 65536%
                },
                xtick style = {draw=none},
                enlarge x limits = 0.075,
                axis y line* = left,
                ylabel style = {align=center},
                ylabel = {\textbf{Avg. Latency} \smaller ( \texttt{ms / batch} )},
                ymin = 0, ymax = 6,
                ytick = {1, 2, 3, 4, 5},
                major tick length=0.3em,
                legend style = {
                    at={(0.62,1.015)}, anchor=south,
                    draw=none,
                    inner sep=0,
                    legend columns=-1,
                    /tikz/every even column/.append style={column sep=2em}
                }
            ]%
                \addlegendentry{T4};
                \addplot[
                    color=LineColorA!50!White,very thick,
                    mark=*,mark options={LineColorA!90!White, scale=0.5}
                ] coordinates {
                    (32,     0.4698455598)[ 0.4698455598]
                    (64,     0.4807016679)[ 0.4807016679]
                    (128,    0.4485380952)[ 0.4485380952]
                    (256,    0.5950644491)[ 0.5950644491]
                    (512,    0.6486721596)[ 0.6486721596]
                    (1024,   0.9242320048)[ 0.9242320048]
                    (2048,   1.640779661 )[ 1.640779661 ]
                    (4096,   2.826187654 )[ 2.826187654 ]
                    (8192,   5.073014553 )[ 5.073014553 ]
                };
                \addlegendentry{A30};
                \addplot[
                    color=LineColorF!70!White,very thick,
                    mark=*,mark options={LineColorF, scale=0.5}
                ] coordinates {
                    (32,     0.3444728916)[ 0.3444728916]
                    (64,     0.6910228311)[ 0.6910228311]
                    (128,    0.6729395712)[ 0.6729395712]
                    (256,    0.6082594059)[ 0.6082594059]
                    (512,    0.6321461412)[ 0.6321461412]
                    (1024,   0.7191909091)[ 0.7191909091]
                    (2048,   1.264726445 )[ 1.264726445 ]
                    (4096,   1.605787879 )[ 1.605787879 ]
                    (8192,   2.72942804  )[ 2.72942804  ]
                };
                \addlegendentry{A100};
                \addplot[
                    color=LineColorC!50!White,very thick,
                    mark=*,mark options={LineColorC!90!White, scale=0.5}
                ] coordinates {
                    (32,     0.2875819113)[ 0.2875819113]
                    (64,     0.3351384892)[ 0.3351384892]
                    (128,    0.4691317158)[ 0.4691317158]
                    (256,    0.5187517986)[ 0.5187517986]
                    (512,    0.5359193878)[ 0.5359193878]
                    (1024,   0.5811115108)[ 0.5811115108]
                    (2048,   0.8193977747)[ 0.8193977747]
                    (4096,   1.161410072 )[ 1.161410072 ]
                    (8192,   1.885827664 )[ 1.885827664 ]
                };
            \end{axis}%
        \end{tikzpicture}}%
        \caption{Criteo 1~TB dataset}%
        \label{f:criteo1tb}%
    \end{subfigure}%
    \hfill%
    \begin{subfigure}[t]{0.32\hsize}%
        \centering%
        \smaller\smaller\smaller%
        \resizebox{\hsize}{!}{\begin{tikzpicture}%
            \begin{axis}[
                width = \figwidth, height = \figheight,
                ybar = 0.45em,
                bar width = 1.1em,
                xlabel = {\textbf{Batch Size}},
                xtick = data, symbolic x coords = {%
                    32, 64, 128, 256, 512, 1024, 2048, 4096, 8192
                },
                xtick style = {draw=none},
                x tick label style={rotate=45,xshift=-0.5em,yshift=0.6em},
                enlarge x limits = 0.075,
                axis y line* = right,
                ylabel style = {align=center, rotate=-180},
                ylabel = {\textbf{Hit rate} \smaller ( \texttt{\%} )},
                ymin = 0.84, ymax = 0.96,
                ytick = {0.87,0.9,0.93},
                yticklabels = {%
                    87,%
                    90,%
                    93,%
                },
                ymajorgrids = true,
                major tick length=0.3em,
                major grid style={thin,dotted,Gray},
                point meta = explicit symbolic,
                legend style = {
                    at={(0.11,1.015)}, anchor=south,
                    draw=none,
                    inner sep=0,
                    legend columns=-1,
                    /tikz/every even column/.append style={column sep=2em}
                },
                legend image code/.code={%
                    \draw[#1,yshift=-0.3em] (0em,0em) rectangle (0.7em,0.7em);
                }
            ]%
                \addlegendentry{Hit rate};
                \addplot[draw=LineColorD,fill=ShadeColorD,text=LineColorD] coordinates {
                    (32,    0.9503113414)[0.9503113414]
                    (64,    0.944796861 )[0.944796861 ]
                    (128,   0.938174762 )[0.938174762 ]
                    (256,   0.9306883521)[0.9306883521]
                    (512,   0.922097759 )[0.922097759 ]
                    (1024,  0.9123444323)[0.9123444323]
                    (2048,  0.9012861126)[0.9012861126]
                    (4096,  0.8887485274)[0.8887485274]
                    (8192,  0.8748170039)[0.8748170039]
                };
            \end{axis}
            \begin{axis}[
                width = \figwidth, height = \figheight,
                axis x line = none,
                xlabel = {\textbf{Batch size}},
                xtick = data, symbolic x coords = {%
                    32, 64, 128, 256, 512, 1024, 2048, 4096, 8192
                },
                xtick style = {draw=none},
                enlarge x limits = 0.075,
                axis y line* = left,
                ylabel style = {align=center},
                ylabel = {\textbf{Avg. Latency} \smaller ( \texttt{ms / batch} )},
                ymin = 0, ymax = 100,
                ytick = {25, 50, 75},
                major tick length=0.3em,
                legend style = {
                    at={(0.62,1.015)}, anchor=south,
                    draw=none,
                    inner sep=0,
                    legend columns=-1,
                    /tikz/every even column/.append style={column sep=2em}
                }
            ]%
                \addlegendentry{T4};
                \addplot[
                    color=LineColorA!50!White,very thick,
                    mark=*,mark options={LineColorA!90!White, scale=0.5}
                ] coordinates {
                    (32,      0.871078)[ 0.871078]
                    (64,      1.080958)[ 1.080958]
                    (128,     2.084895)[ 2.084895]
                    (256,     3.091692)[ 3.091692]
                    (512,     9.146828)[ 9.146828]
                    (1024,   13.048576)[13.048576]
                    (2048,   25.221714)[25.221714]
                    (4096,   42.71374 )[42.71374 ]
                    (8192,   88.425117)[88.425117]
                };
                \addlegendentry{A30};
                \addplot[
                    color=LineColorF!70!White,very thick,
                    mark=*,mark options={LineColorF, scale=0.5}
                ] coordinates {
                    (32,      0.721225058)[  0.721225058]
                    (64,      0.891025355)[  0.891025355]
                    (128,     1.342232456)[  1.342232456]
                    (256,     2.385526316)[  2.385526316]
                    (512,     4.1079839  )[  4.1079839  ]
                    (1024,    7.404429032)[  7.404429032]
                    (2048,   14.94513031 )[ 14.94513031 ]
                    (4096,   27.31018078 )[ 27.31018078 ]
                    (8192,   53.99418906 )[ 53.99418906 ]
                };
                \addlegendentry{A100};
                \addplot[
                    color=LineColorC!50!White,very thick,
                    mark=*,mark options={LineColorC!90!White, scale=0.5}
                ] coordinates {
                    (32,      0.6092552188)[  0.6092552188]
                    (64,      0.7407793728)[  0.7407793728]
                    (128,     1.034970386 )[  1.034970386 ]
                    (256,     1.590935848 )[  1.590935848 ]
                    (512,     2.802354975 )[  2.802354975 ]
                    (1024,    4.622808321 )[  4.622808321 ]
                    (2048,    8.633312714 )[  8.633312714 ]
                    (4096,   16.26296615  )[ 16.26296615  ]
                    (8192,   31.39529114  )[ 31.39529114  ]
                };
            \end{axis}%
        \end{tikzpicture}}%
        \caption{Synthetic dataset B}%
        \label{f:new_synthetic}%
    \end{subfigure}%
    
    \caption{Comparison of HPS performance with different NVIDIA GPUs.}
    \label{f:inf-latency-comparison}
\end{figure*}
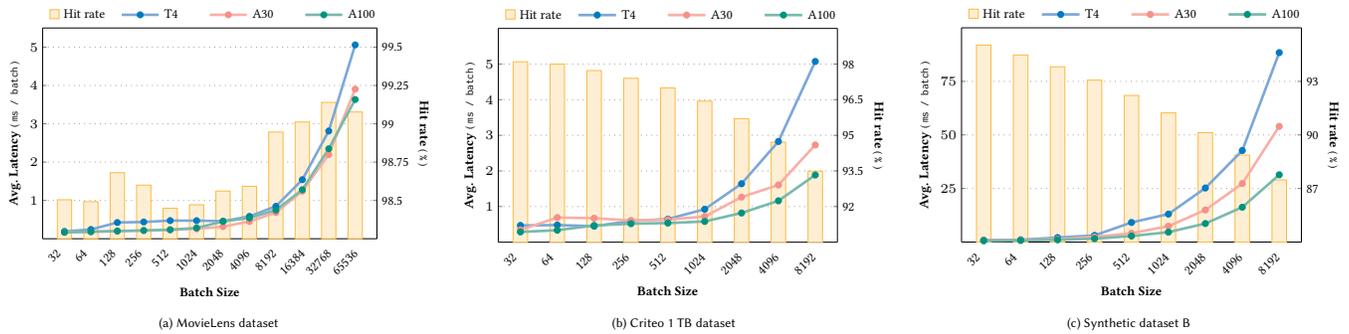

\subsubsection{Performance and accuracy comparison}\label{perf_comp}
To determine the influence of the hardware on HPS performance, we measure the stable stage inference latency with different datasets and batch sizes on respectively a NVIDIA T4 (16~GB memory), A30 (24~GB memory) and A100 GPU (80~GB memory). To allow for a fair comparison despite the limited memory compliment of the T4 GPU, we set the cache percentage and hit rate threshold to respectively 10\% and 1.0 for all GPUs. Hence, for the same dataset, the different GPUs stabilize at the same hit rate throughout \autoref{f:inf-latency-comparison}. For the comparatively tiny \emph{MovieLens} (\autoref{f:movielens}) dataset, we can achieve an inference latency of less than 1~\emph{ms} with the hit rate saturating at \textasciitilde98.5\%. With \emph{Criteo 1~TB} (\autoref{f:criteo1tb}) and \emph{Synthetic dataset B} (\autoref{f:new_synthetic}), the hit rate gradually decreases as the batch size increases. \emph{Synthetic dataset B} simulates a recommendation task where more categorical features are used. Larger batch sizes lead to increased computational overhead of the dense layers and higher overall inference latency. Thus, although their embedding tables are similarly sized, the latency figures for \emph{Synthetic dataset B} are much higher than those for the \emph{Criteo 1~TB} dataset. 
Because large recommendation models often do not fit into GPU memory, most inference frameworks only load the dense model part into GPU memory. Meanwhile, HPS can deploy and accelerate such models on the GPU through its GPU cache mechanism. When using small and medium batch sizes with HPS, the inference latencies for T4 and A30 GPUs are on par with an A100 GPU, thus demonstrating that HPS is a scalable inference solution for recommender systems. To complete our investigation, we present the prediction accuracy, \emph{i.e.}, $\frac{\sum \text{correct}}{\sum \text{total}}$ samples, when varying the cache hit rate for the \emph{Criteo 1~TB} dataset in \autoref{f:acc_hit_rate}. Here, the three different hit rate thresholds overlap almost perfectly, while their cache hit rates stabilize above 0.9. This strongly implies that the GPU embedding cache retains hot embeddings well, even if the asynchronous insertion is employed.

\subsection{Online update performance}\label{s:impact-online-updates}
Our updating mechanism consists of three major components (\emph{cf.} \autoref{s:update}): (1) dumping the model in the training nodes, (2) update ingestion by the inference nodes, and (3) the embedding cache refresh operation. Because model dumps are done in isolation from the overall HPS inference system, we only focus on the remaining two components.

\textbf{Update ingestion mechanism}. The delivery ratio of model parameter updates depends heavily on the configuration of the intermediate Kafka message buffer storage and its network connectivity. Of the latter, we have plenty (\emph{cf.} \autoref{s:cluster-setup}). Therefore, we concentrate on the receiving VDB/PDB instances. In \autoref{t:database-backend-insert-perf}, we perform asynchronous random batch insertion (batch size = 128 MB) limits for \emph{Synthetic dataset A} in our test environment. The insertion speed slowly declines as the model size increases due to storage management overheads.

\textbf{GPU embedding cache refresh.} In \autoref{t:ec-refresh-table}, we analyze the embedding cache refresh performance with different cache capacities, and show the impact of latency and capacity. Note how the overhead for actually \emph{dumping} the embedding keys scales with the cache size, but is almost negligible in comparison to the following \emph{update} operation. The throughput remains stable around 199 GB/s, so that it only takes 200 ms to refresh a 40~GB embedding cache.

To summarize, online updates only have a minor impact on the overall inference performance. That is because the VDB/PDB update operations happen lazily and infrequently, and subsequent cache refreshes happen near-instantaneously.

\section{Conclusion}\label{s:conclusion}

In this paper we presented and analyzed HPS, an efficient GPU-enabled hierarchical parameter server for building large-scale model inference services. Our high-performance GPU embedding cache exploits the typical properties of recommendation datasets to improve inference throughput. By extending this GPU embedding cache with other cluster storage resources (VDB \& PDB), HPS can efficiently process queries for very large models. Through its asynchronous update mechanisms, HPS ensures that its GPU embedding cache retains a high hit rate over time. 

Our experiments show that the HugeCTR HPS can reduce the latency for end-to-end model inference by 5-62x in comparison with PyTorch CPU. Furthermore, HPS offers excellent scaling and performance on different GPUs. 

As for future work, we intend to continue extending Merlin HugeCTR and HPS with additional features, including but not limited to better support for next generation GPU technologies and further optimizing performance, like relaxing the constraints when using locks to protect data and ensure thread-safety for embedding cache.

\bibliographystyle{ACM-Reference-Format}
\bibliography{references}


\begin{thebibliography}{42}


\ifx \showCODEN    \undefined \def \showCODEN     #1{\unskip}     \fi
\ifx \showDOI      \undefined \def \showDOI       #1{#1}\fi
\ifx \showISBNx    \undefined \def \showISBNx     #1{\unskip}     \fi
\ifx \showISBNxiii \undefined \def \showISBNxiii  #1{\unskip}     \fi
\ifx \showISSN     \undefined \def \showISSN      #1{\unskip}     \fi
\ifx \showLCCN     \undefined \def \showLCCN      #1{\unskip}     \fi
\ifx \shownote     \undefined \def \shownote      #1{#1}          \fi
\ifx \showarticletitle \undefined \def \showarticletitle #1{#1}   \fi
\ifx \showURL      \undefined \def \showURL       {\relax}        \fi
\providecommand\bibfield[2]{#2}
\providecommand\bibinfo[2]{#2}
\providecommand\natexlab[1]{#1}
\providecommand\showeprint[2][]{arXiv:#2}

\bibitem[Arefyeva et~al\mbox{.}(2018)]%
        {arefyeva2018cpu-gpu-db-survey}
\bibfield{author}{\bibinfo{person}{Iya Arefyeva}, \bibinfo{person}{David
  Broneske}, \bibinfo{person}{Gabriel Campero}, \bibinfo{person}{Marcus
  Pinnecke}, {and} \bibinfo{person}{Gunter Saake}.}
  \bibinfo{year}{2018}\natexlab{}.
\newblock \showarticletitle{{Memory Management Strategies in CPU/GPU Database
  Systems: A Survey}}. In \bibinfo{booktitle}{\emph{Beyond Databases,
  Architectures and Structures. Facing the Challenges of Data Proliferation and
  Growing Variety}}. \bibinfo{publisher}{Springer International Publishing},
  \bibinfo{address}{Cham}, \bibinfo{pages}{128--142}.
\newblock
\showISBNx{978-3-319-99987-6}


\bibitem[Ben-Nun and Hoefler(2019)]%
        {bennun2019dist-deep-learning}
\bibfield{author}{\bibinfo{person}{Tal Ben-Nun} {and} \bibinfo{person}{Torsten
  Hoefler}.} \bibinfo{year}{2019}\natexlab{}.
\newblock \showarticletitle{{Demystifying Parallel and Distributed Deep
  Learning: An In-Depth Concurrency Analysis}}.
\newblock \bibinfo{journal}{\emph{ACM Comput. Surv.}} \bibinfo{volume}{52},
  \bibinfo{number}{4}, Article \bibinfo{articleno}{65} (\bibinfo{date}{aug}
  \bibinfo{year}{2019}), \bibinfo{numpages}{43}~pages.
\newblock
\showISSN{0360-0300}
\urldef\tempurl%
\url{https://doi.org/10.1145/3320060}
\showDOI{\tempurl}


\bibitem[Clauset et~al\mbox{.}(2009)]%
        {clauset2009power-law-dist}
\bibfield{author}{\bibinfo{person}{Aaron Clauset},
  \bibinfo{person}{Cosma~Rohilla Shalizi}, {and} \bibinfo{person}{M.~E.~J.
  Newman}.} \bibinfo{year}{2009}\natexlab{}.
\newblock \showarticletitle{{Power-Law Distributions in Empirical Data}}.
\newblock \bibinfo{journal}{\emph{SIAM Rev.}} \bibinfo{volume}{51},
  \bibinfo{number}{4} (\bibinfo{year}{2009}), \bibinfo{pages}{661--703}.
\newblock
\showISSN{0036-1445}
\urldef\tempurl%
\url{https://doi.org/10.1137/070710111}
\showDOI{\tempurl}


\bibitem[Collet(2014)]%
        {collet2014xxhash}
\bibfield{author}{\bibinfo{person}{Yann Collet}.}
  \bibinfo{year}{2014}\natexlab{}.
\newblock \bibinfo{title}{{xxHash Hash Function}}.
\newblock \bibinfo{howpublished}{\url{https://www.xxhash.com}}.
\newblock
\newblock
\shownote{Accessed: 2022-04-15}.


\bibitem[Crankshaw et~al\mbox{.}(2017)]%
        {Clipper}
\bibfield{author}{\bibinfo{person}{Daniel Crankshaw}, \bibinfo{person}{Xin
  Wang}, \bibinfo{person}{Guilio Zhou}, \bibinfo{person}{Michael~J. Franklin},
  \bibinfo{person}{Joseph~E. Gonzalez}, {and} \bibinfo{person}{Ion Stoica}.}
  \bibinfo{year}{2017}\natexlab{}.
\newblock \showarticletitle{{Clipper: A Low-Latency Online Prediction Serving
  System}}. In \bibinfo{booktitle}{\emph{14th USENIX Symposium on Networked
  Systems Design and Implementation (NSDI 17)}}. \bibinfo{publisher}{USENIX
  Association}, \bibinfo{address}{Boston, MA}, \bibinfo{pages}{613--627}.
\newblock
\showISBNx{978-1-931971-37-9}


\bibitem[{Criteo AI Lab}(2014)]%
        {criteo2014dataset}
\bibfield{author}{\bibinfo{person}{{Criteo AI Lab}}.}
  \bibinfo{year}{2014}\natexlab{}.
\newblock \bibinfo{title}{{Criteo 1TB Click Logs dataset}}.
\newblock
  \bibinfo{howpublished}{\url{https://www.kaggle.com/c/criteo-display-ad-challenge}}.
\newblock
\newblock
\shownote{Accessed: 2022-03-15}.


\bibitem[Cui et~al\mbox{.}(2016)]%
        {GeePS}
\bibfield{author}{\bibinfo{person}{Henggang Cui}, \bibinfo{person}{Hao Zhang},
  \bibinfo{person}{Gregory~R. Ganger}, \bibinfo{person}{Phillip~B. Gibbons},
  {and} \bibinfo{person}{Eric~P. Xing}.} \bibinfo{year}{2016}\natexlab{}.
\newblock \showarticletitle{{GeePS: Scalable Deep Learning on Distributed GPUs
  with a GPU-Specialized Parameter Server}}. In
  \bibinfo{booktitle}{\emph{Proceedings of the 11th European Conference on
  Computer Systems}} (London, United Kingdom) \emph{(\bibinfo{series}{EuroSys
  '16})}. \bibinfo{publisher}{Association for Computing Machinery},
  \bibinfo{address}{New York, NY, USA}, Article \bibinfo{articleno}{4},
  \bibinfo{numpages}{16}~pages.
\newblock
\showISBNx{9781450342407}
\urldef\tempurl%
\url{https://doi.org/10.1145/2901318.2901323}
\showDOI{\tempurl}


\bibitem[Farrell et~al\mbox{.}(2021)]%
        {farrell2021mlperf}
\bibfield{author}{\bibinfo{person}{Steven Farrell}, \bibinfo{person}{Murali
  Emani}, \bibinfo{person}{Jacob Balma}, \bibinfo{person}{Lukas Drescher},
  \bibinfo{person}{Aleksandr Drozd}, \bibinfo{person}{Andreas Fink},
  \bibinfo{person}{Geoffrey Fox}, \bibinfo{person}{David Kanter},
  \bibinfo{person}{Thorsten Kurth}, \bibinfo{person}{Peter Mattson},
  \bibinfo{person}{Dawei Mu}, \bibinfo{person}{Amit Ruhela},
  \bibinfo{person}{Kento Sato}, \bibinfo{person}{Koichi Shirahata},
  \bibinfo{person}{Tsuguchika Tabaru}, \bibinfo{person}{Aristeidis Tsaris},
  \bibinfo{person}{Jan Balewski}, \bibinfo{person}{Ben Cumming},
  \bibinfo{person}{Takumi Danjo}, \bibinfo{person}{Jens Domke},
  \bibinfo{person}{Takaaki Fukai}, \bibinfo{person}{Naoto Fukumoto},
  \bibinfo{person}{Tatsuya Fukushi}, \bibinfo{person}{Balazs Gerofi},
  \bibinfo{person}{Takumi Honda}, \bibinfo{person}{Toshiyuki Imamura},
  \bibinfo{person}{Akihiko Kasagi}, \bibinfo{person}{Kentaro Kawakami},
  \bibinfo{person}{Shuhei Kudo}, \bibinfo{person}{Akiyoshi Kuroda},
  \bibinfo{person}{Maxime Martinasso}, \bibinfo{person}{Satoshi Matsuoka},
  \bibinfo{person}{Henrique Mendonça}, \bibinfo{person}{Kazuki Minami},
  \bibinfo{person}{Prabhat Ram}, \bibinfo{person}{Takashi Sawada},
  \bibinfo{person}{Mallikarjun Shankar}, \bibinfo{person}{Tom~St. John},
  \bibinfo{person}{Akihiro Tabuchi}, \bibinfo{person}{Venkatram Vishwanath},
  \bibinfo{person}{Mohamed Wahib}, \bibinfo{person}{Masafumi Yamazaki}, {and}
  \bibinfo{person}{Junqi Yin}.} \bibinfo{year}{2021}\natexlab{}.
\newblock \showarticletitle{{MLPerf\texttrademark{} HPC: A Holistic Benchmark
  Suite for Scientific Machine Learning on HPC Systems}}. In
  \bibinfo{booktitle}{\emph{IEEE/ACM Workshop on Machine Learning in High
  Performance Computing Environments (MLHPC)}} (St. Louis, MO, USA).
  \bibinfo{publisher}{IEEE Press}, \bibinfo{address}{New York, NY, USA},
  \bibinfo{pages}{33--45}.
\newblock
\showISBNx{978-1-6654-1125-7}
\showISSN{2768-4253}
\urldef\tempurl%
\url{https://doi.org/10.1109/MLHPC54614.2021.00009}
\showDOI{\tempurl}


\bibitem[Goodwin et~al\mbox{.}(2021a)]%
        {triton_model_management}
\bibfield{author}{\bibinfo{person}{David Goodwin} {et~al\mbox{.}}}
  \bibinfo{year}{2021}\natexlab{a}.
\newblock \bibinfo{title}{{NVIDIA Triton: Model Control Mode}}.
\newblock
  \bibinfo{howpublished}{\url{https://github.com/triton-inference-server/server/blob/main/docs/model_management.md\#model-control-mode-explicit}}.
\newblock
\newblock
\shownote{Accessed: 2022-07-19}.


\bibitem[Goodwin et~al\mbox{.}(2021b)]%
        {triton_instance_group}
\bibfield{author}{\bibinfo{person}{David Goodwin} {et~al\mbox{.}}}
  \bibinfo{year}{2021}\natexlab{b}.
\newblock \bibinfo{title}{{Triton Architecture - Concurrent Model Execution}}.
\newblock
  \bibinfo{howpublished}{\url{https://github.com/triton-inference-server/server/blob/main/docs/architecture.md\#concurrent-model-execution}}.
\newblock
\newblock
\shownote{Accessed: 2022-07-19}.


\bibitem[Goodwin et~al\mbox{.}(2022)]%
        {triton_perf_analyzer}
\bibfield{author}{\bibinfo{person}{David Goodwin} {et~al\mbox{.}}}
  \bibinfo{year}{2022}\natexlab{}.
\newblock \bibinfo{title}{{NVIDIA Triton: Performance Analyzer}}.
\newblock
  \bibinfo{howpublished}{\url{https://github.com/triton-inference-server/server/blob/main/docs/perf_analyzer.md}}.
\newblock
\newblock
\shownote{Accessed: 2022-07-19}.


\bibitem[Guo et~al\mbox{.}(2021)]%
        {ScaleFreeCTR}
\bibfield{author}{\bibinfo{person}{Huifeng Guo}, \bibinfo{person}{Wei Guo},
  \bibinfo{person}{Yong Gao}, \bibinfo{person}{Ruiming Tang},
  \bibinfo{person}{Xiuqiang He}, {and} \bibinfo{person}{Wenzhi Liu}.}
  \bibinfo{year}{2021}\natexlab{}.
\newblock \bibinfo{booktitle}{\emph{{ScaleFreeCTR: MixCache-Based Distributed
  Training System for CTR Models with Huge Embedding Table}}}.
\newblock \bibinfo{publisher}{Association for Computing Machinery},
  \bibinfo{address}{New York, NY, USA}, \bibinfo{pages}{1269--1278}.
\newblock
\showISBNx{9781450380379}
\urldef\tempurl%
\url{https://doi.org/10.1145/3404835.3462976}
\showDOI{\tempurl}


\bibitem[Gupta et~al\mbox{.}(2020)]%
        {DeepRecSys}
\bibfield{author}{\bibinfo{person}{Udit Gupta}, \bibinfo{person}{Samuel Hsia},
  \bibinfo{person}{Vikram Saraph}, \bibinfo{person}{Xiaodong Wang},
  \bibinfo{person}{Brandon Reagen}, \bibinfo{person}{Gu-Yeon Wei},
  \bibinfo{person}{Hsien-Hsin~S. Lee}, \bibinfo{person}{David Brooks}, {and}
  \bibinfo{person}{Carole-Jean Wu}.} \bibinfo{year}{2020}\natexlab{}.
\newblock \showarticletitle{{DeepRecSys: A System for Optimizing End-To-End
  At-Scale Neural Recommendation Inference}}. In
  \bibinfo{booktitle}{\emph{ACM/IEEE 47th Annual International Symposium on
  Computer Architecture (ISCA)}}. \bibinfo{publisher}{IEEE Press},
  \bibinfo{address}{Valencia, Spain}, \bibinfo{pages}{982--995}.
\newblock
\urldef\tempurl%
\url{https://doi.org/10.1109/ISCA45697.2020.00084}
\showDOI{\tempurl}


\bibitem[Gupta et~al\mbox{.}(2021)]%
        {Fast_Distributed_Training}
\bibfield{author}{\bibinfo{person}{Vipul Gupta}, \bibinfo{person}{Dhruv
  Choudhary}, \bibinfo{person}{Peter Tang}, \bibinfo{person}{Xiaohan Wei},
  \bibinfo{person}{Xing Wang}, \bibinfo{person}{Yuzhen Huang},
  \bibinfo{person}{Arun Kejariwal}, \bibinfo{person}{Kannan Ramchandran}, {and}
  \bibinfo{person}{Michael~W. Mahoney}.} \bibinfo{year}{2021}\natexlab{}.
\newblock \showarticletitle{{Training Recommender Systems at Scale:
  Communication-Efficient Model and Data Parallelism}}. In
  \bibinfo{booktitle}{\emph{Proceedings of the 27th ACM SIGKDD Conference on
  Knowledge Discovery \& Data Mining}} (Virtual Event, Singapore)
  \emph{(\bibinfo{series}{KDD '21})}. \bibinfo{publisher}{Association for
  Computing Machinery}, \bibinfo{address}{New York, NY, USA},
  \bibinfo{pages}{2928--2936}.
\newblock
\showISBNx{9781450383325}
\urldef\tempurl%
\url{https://doi.org/10.1145/3447548.3467080}
\showDOI{\tempurl}


\bibitem[Harper and Konstan(2015)]%
        {movielens-25m}
\bibfield{author}{\bibinfo{person}{F.~Maxwell Harper} {and}
  \bibinfo{person}{Joseph~A. Konstan}.} \bibinfo{year}{2015}\natexlab{}.
\newblock \showarticletitle{The MovieLens Datasets: History and Context}.
\newblock \bibinfo{journal}{\emph{ACM Trans. Interact. Intell. Syst.}}
  \bibinfo{volume}{5}, \bibinfo{number}{4}, Article \bibinfo{articleno}{19}
  (\bibinfo{year}{2015}), \bibinfo{numpages}{19}~pages.
\newblock
\showISSN{2160-6455}
\urldef\tempurl%
\url{https://doi.org/10.1145/2827872}
\showDOI{\tempurl}


\bibitem[Huang et~al\mbox{.}(2018)]%
        {FlexPS}
\bibfield{author}{\bibinfo{person}{Yuzhen Huang}, \bibinfo{person}{Tatiana
  Jin}, \bibinfo{person}{Yidi Wu}, \bibinfo{person}{Zhenkun Cai},
  \bibinfo{person}{Xiao Yan}, \bibinfo{person}{Fan Yang},
  \bibinfo{person}{Jinfeng Li}, \bibinfo{person}{Yuying Guo}, {and}
  \bibinfo{person}{James Cheng}.} \bibinfo{year}{2018}\natexlab{}.
\newblock \showarticletitle{{FlexPS: Flexible Parallelism Control in Parameter
  Server Architecture}}.
\newblock \bibinfo{journal}{\emph{Proc. VLDB Endow.}} \bibinfo{volume}{11},
  \bibinfo{number}{5} (\bibinfo{year}{2018}), \bibinfo{pages}{566--579}.
\newblock
\showISSN{2150-8097}
\urldef\tempurl%
\url{https://doi.org/10.1145/3177732.3177734}
\showDOI{\tempurl}


\bibitem[Jiang et~al\mbox{.}(2017)]%
        {Angel}
\bibfield{author}{\bibinfo{person}{Jie Jiang}, \bibinfo{person}{Lele Yu},
  \bibinfo{person}{Jiawei Jiang}, \bibinfo{person}{Yuhong Liu}, {and}
  \bibinfo{person}{Bin Cui}.} \bibinfo{year}{2017}\natexlab{}.
\newblock \showarticletitle{Angel: a new large-scale machine learning system}.
\newblock \bibinfo{journal}{\emph{National Science Review}}
  \bibinfo{volume}{5}, \bibinfo{number}{2} (\bibinfo{year}{2017}),
  \bibinfo{pages}{216--236}.
\newblock
\showISSN{2095-5138}
\urldef\tempurl%
\url{https://doi.org/10.1093/nsr/nwx018}
\showDOI{\tempurl}


\bibitem[Jiang et~al\mbox{.}(2021)]%
        {jiang2021fleet-rec}
\bibfield{author}{\bibinfo{person}{Wenqi Jiang}, \bibinfo{person}{Zhenhao He},
  \bibinfo{person}{Shuai Zhang}, \bibinfo{person}{Kai Zeng},
  \bibinfo{person}{Liang Feng}, \bibinfo{person}{Jiansong Zhang},
  \bibinfo{person}{Tongxuan Liu}, \bibinfo{person}{Yong Li},
  \bibinfo{person}{Jingren Zhou}, \bibinfo{person}{Ce Zhang}, {and}
  \bibinfo{person}{Gustavo Alonso}.} \bibinfo{year}{2021}\natexlab{}.
\newblock \showarticletitle{FleetRec: Large-Scale Recommendation Inference on
  Hybrid GPU-FPGA Clusters}. In \bibinfo{booktitle}{\emph{Proceedings of the
  27th ACM SIGKDD Conference on Knowledge Discovery \& Data Mining}} (Virtual
  Event, Singapore) \emph{(\bibinfo{series}{KDD '21})}.
  \bibinfo{publisher}{Association for Computing Machinery},
  \bibinfo{address}{New York, NY, USA}, \bibinfo{pages}{3097--3105}.
\newblock
\showISBNx{9781450383325}
\urldef\tempurl%
\url{https://doi.org/10.1145/3447548.3467139}
\showDOI{\tempurl}


\bibitem[Kanter et~al\mbox{.}(2021)]%
        {kanter2021mlperf-1-1-results}
\bibfield{author}{\bibinfo{person}{David Kanter}, \bibinfo{person}{Peter
  Mattson}, {et~al\mbox{.}}} \bibinfo{year}{2021}\natexlab{}.
\newblock \bibinfo{title}{{ML$\cdot$Commons / MLperf v1.1 Results}}.
\newblock
  \bibinfo{howpublished}{\url{https://mlcommons.org/en/training-normal-11}}.
\newblock
\newblock
\shownote{Accessed: 2022-03-15}.


\bibitem[Langer et~al\mbox{.}(2020)]%
        {langer2020ddls}
\bibfield{author}{\bibinfo{person}{Matthias Langer}, \bibinfo{person}{Zhen He},
  \bibinfo{person}{Yanbo Xue}, {and} \bibinfo{person}{Wenny Rahayu}.}
  \bibinfo{year}{2020}\natexlab{}.
\newblock \showarticletitle{{Distributed Training of Deep Learning Models: A
  Taxonomic Perspective}}.
\newblock \bibinfo{journal}{\emph{IEEE Transactions on Parallel and Distributed
  Systems}} \bibinfo{volume}{31}, \bibinfo{number}{12} (\bibinfo{year}{2020}),
  \bibinfo{pages}{2802--2818}.
\newblock
\showISSN{2331-8422}
\urldef\tempurl%
\url{https://doi.org/10.1109/TPDS.2020.3003307}
\showDOI{\tempurl}


\bibitem[Lui et~al\mbox{.}(2021)]%
        {Capacity-Driven}
\bibfield{author}{\bibinfo{person}{Michael Lui}, \bibinfo{person}{Yavuz Yetim},
  \bibinfo{person}{{\"{O}}zg{\"{u}}r {\"{O}}zkan}, \bibinfo{person}{Zhuoran
  Zhao}, \bibinfo{person}{Shin-Yeh Tsai}, \bibinfo{person}{Carole-Jean Wu},
  {and} \bibinfo{person}{Mark Hempstead}.} \bibinfo{year}{2021}\natexlab{}.
\newblock \showarticletitle{{Understanding Capacity-Driven Scale-Out Neural
  Recommendation Inference}}. In \bibinfo{booktitle}{\emph{2021 IEEE
  International Symposium on Performance Analysis of Systems and Software
  (ISPASS)}} (Stony Brook, NY, USA). \bibinfo{publisher}{IEEE Press},
  \bibinfo{address}{New York, NY, USA}, \bibinfo{pages}{162--171}.
\newblock
\urldef\tempurl%
\url{https://doi.org/10.1109/ISPASS51385.2021.00033}
\showDOI{\tempurl}


\bibitem[Luo et~al\mbox{.}(2018)]%
        {Parameter_Hub}
\bibfield{author}{\bibinfo{person}{Liang Luo}, \bibinfo{person}{Jacob Nelson},
  \bibinfo{person}{Luis Ceze}, \bibinfo{person}{Amar Phanishayee}, {and}
  \bibinfo{person}{Arvind Krishnamurthy}.} \bibinfo{year}{2018}\natexlab{}.
\newblock \showarticletitle{{Parameter Hub: A Rack-Scale Parameter Server for
  Distributed Deep Neural Network Training}}. In
  \bibinfo{booktitle}{\emph{Proceedings of the ACM Symposium on Cloud
  Computing}} (Carlsbad, CA, USA) \emph{(\bibinfo{series}{SoCC '18})}.
  \bibinfo{publisher}{Association for Computing Machinery},
  \bibinfo{address}{New York, NY, USA}, \bibinfo{pages}{41--54}.
\newblock
\showISBNx{9781450360111}
\urldef\tempurl%
\url{https://doi.org/10.1145/3267809.3267840}
\showDOI{\tempurl}


\bibitem[Mittal and Vetter(2015)]%
        {mittal2015cpu-gpu-heterogeneous-survey}
\bibfield{author}{\bibinfo{person}{Sparsh Mittal} {and}
  \bibinfo{person}{Jeffrey~S. Vetter}.} \bibinfo{year}{2015}\natexlab{}.
\newblock \showarticletitle{{A Survey of CPU-GPU Heterogeneous Computing
  Techniques}}.
\newblock \bibinfo{journal}{\emph{ACM Comput. Surv.}} \bibinfo{volume}{47},
  \bibinfo{number}{4}, Article \bibinfo{articleno}{69} (\bibinfo{year}{2015}),
  \bibinfo{numpages}{35}~pages.
\newblock
\showISSN{0360-0300}
\urldef\tempurl%
\url{https://doi.org/10.1145/2788396}
\showDOI{\tempurl}


\bibitem[Naumov et~al\mbox{.}(2019)]%
        {dlrm}
\bibfield{author}{\bibinfo{person}{Maxim Naumov}, \bibinfo{person}{Dheevatsa
  Mudigere}, \bibinfo{person}{Hao-Jun~Michael Shi}, \bibinfo{person}{Jianyu
  Huang}, \bibinfo{person}{Narayanan Sundaraman}, \bibinfo{person}{Jongsoo
  Park}, \bibinfo{person}{Xiaodong Wang}, \bibinfo{person}{Udit Gupta},
  \bibinfo{person}{Carole-Jean Wu}, \bibinfo{person}{Alisson~G. Azzolini},
  \bibinfo{person}{Dmytro Dzhulgakov}, \bibinfo{person}{Andrey Mallevich},
  \bibinfo{person}{Ilia Cherniavskii}, \bibinfo{person}{Yinghai Lu},
  \bibinfo{person}{Raghuraman Krishnamoorthi}, \bibinfo{person}{Ansha Yu},
  \bibinfo{person}{Volodymyr Kondratenko}, \bibinfo{person}{Stephanie Pereira},
  \bibinfo{person}{Xianjie Chen}, \bibinfo{person}{Wenlin Chen},
  \bibinfo{person}{Vijay Rao}, \bibinfo{person}{Bill Jia},
  \bibinfo{person}{Liang Xiong}, {and} \bibinfo{person}{Misha Smelyanskiy}.}
  \bibinfo{year}{2019}\natexlab{}.
\newblock \showarticletitle{{Deep Learning Recommendation Model for
  Personalization and Recommendation Systems}}.
\newblock \bibinfo{journal}{\emph{CoRR}}  \bibinfo{volume}{abs/1906.00091}
  (\bibinfo{year}{2019}), \bibinfo{numpages}{10}~pages.
\newblock
\urldef\tempurl%
\url{http://arxiv.org/abs/1906.00091}
\showURL{%
\tempurl}


\bibitem[NVIDA(2022a)]%
        {nvidia2022hps-dist-deployment}
\bibfield{author}{\bibinfo{person}{NVIDA}.} \bibinfo{year}{2022}\natexlab{a}.
\newblock \bibinfo{title}{{HugeCTR: Distributed deployment with the
  Hierarchical Parameter Server}}.
\newblock
  \bibinfo{howpublished}{\url{https://github.com/triton-inference-server/hugectr_backend/blob/main/docs/architecture.md\#distributed-deployment-with-hierarchical-hugectr-parameter-server}}.
\newblock
\newblock
\shownote{Accessed: 2022-04-15}.


\bibitem[NVIDA(2022b)]%
        {nvidia2022hps}
\bibfield{author}{\bibinfo{person}{NVIDA}.} \bibinfo{year}{2022}\natexlab{b}.
\newblock \bibinfo{title}{{HugeCTR Hierarchical Parameter Server}}.
\newblock
  \bibinfo{howpublished}{\url{https://github.com/triton-inference-server/hugectr_backend/blob/main/docs/hierarchical_parameter_server.md\#hugectr-hierarchical-parameter-server}}.
\newblock
\newblock
\shownote{Accessed: 2022-04-15}.


\bibitem[NVIDIA(2020)]%
        {nvidia2020ampere-arch}
\bibfield{author}{\bibinfo{person}{NVIDIA}.} \bibinfo{year}{2020}\natexlab{}.
\newblock \bibinfo{title}{{NVIDIA A100 Tensor Core GPU}}.
\newblock
  \bibinfo{howpublished}{\url{https://images.nvidia.com/aem-dam/en-zz/Solutions/data-center/nvidia-ampere-architecture-whitepaper.pdf}}.
  , \bibinfo{numpages}{82}~pages.
\newblock


\bibitem[NVIDIA(2022a)]%
        {CUDA_Warp}
\bibfield{author}{\bibinfo{person}{NVIDIA}.} \bibinfo{year}{2022}\natexlab{a}.
\newblock \bibinfo{title}{{CUDA C Programming Guide}}.
\newblock
  \bibinfo{howpublished}{\url{https://docs.nvidia.com/cuda/cuda-c-programming-guide/index.html\#simt-architecture}}.
  , \bibinfo{numpages}{131}~pages.
\newblock
\newblock
\shownote{Accessed: 2022-04-15}.


\bibitem[NVIDIA(2022b)]%
        {nvidia2022hps-async-data-insert}
\bibfield{author}{\bibinfo{person}{NVIDIA}.} \bibinfo{year}{2022}\natexlab{b}.
\newblock \bibinfo{title}{{HugeCTR: Embedding Cache Asynchronous Insertion
  Mechanism}}.
\newblock
  \bibinfo{howpublished}{\url{https://github.com/triton-inference-server/hugectr_backend\#embedding-cache-asynchronous-insertion-mechanism}}.
\newblock
\newblock
\shownote{Accessed: 2022-04-15}.


\bibitem[NVIDIA(2022c)]%
        {nvidia2022hps-gpu-cache}
\bibfield{author}{\bibinfo{person}{NVIDIA}.} \bibinfo{year}{2022}\natexlab{c}.
\newblock \bibinfo{title}{{HugeCTR: GPU Cache}}.
\newblock
  \bibinfo{howpublished}{\url{https://github.com/NVIDIA-Merlin/HugeCTR/tree/master/gpu_cache}}.
\newblock
\newblock
\shownote{Accessed: 2022-04-15}.


\bibitem[NVIDIA(2022d)]%
        {nvidia2022hugectr-backend}
\bibfield{author}{\bibinfo{person}{NVIDIA}.} \bibinfo{year}{2022}\natexlab{d}.
\newblock \bibinfo{title}{{HugeCTR: Triton Backend}}.
\newblock
  \bibinfo{howpublished}{\url{https://github.com/triton-inference-server/hugectr_backend}}.
\newblock
\newblock
\shownote{Accessed: 2022-05-15}.


\bibitem[{NVidia Corp.}(2022)]%
        {nvidia2022data-center}
\bibfield{author}{\bibinfo{person}{{NVidia Corp.}}}
  \bibinfo{year}{2022}\natexlab{}.
\newblock \bibinfo{title}{{Solutions for the Data-Center}}.
\newblock
  \bibinfo{howpublished}{\url{https://www.nvidia.com/en-us/data-center}}.
\newblock
\newblock
\shownote{Accessed: 2022-04-15}.


\bibitem[{NVIDIA Deep Learning Examples for Tensor Cores}(2022a)]%
        {torch-infer-cpu}
\bibfield{author}{\bibinfo{person}{{NVIDIA Deep Learning Examples for Tensor
  Cores}}.} \bibinfo{year}{2022}\natexlab{a}.
\newblock \bibinfo{title}{{Deploying the DLRM model using Triton Inference
  Server}}.
\newblock
  \bibinfo{howpublished}{\url{https://github.com/NVIDIA/DeepLearningExamples/tree/master/PyTorch/Recommendation/DLRM/triton\#performance}}.
\newblock


\bibitem[{NVIDIA Deep Learning Examples for Tensor Cores}(2022b)]%
        {tf-infer-gpu}
\bibfield{author}{\bibinfo{person}{{NVIDIA Deep Learning Examples for Tensor
  Cores}}.} \bibinfo{year}{2022}\natexlab{b}.
\newblock \bibinfo{title}{{DLRM For TensorFlow 2}}.
\newblock
  \bibinfo{howpublished}{\url{https://github.com/NVIDIA/DeepLearningExamples/tree/master/TensorFlow2/Recommendation/DLRM\#inference-performance-results}}.
\newblock


\bibitem[Rosenfeld et~al\mbox{.}(2022)]%
        {rosenfeld2022cpu-gpu-query-processing-survey}
\bibfield{author}{\bibinfo{person}{Viktor Rosenfeld},
  \bibinfo{person}{Sebastian Bre\ss{}}, {and} \bibinfo{person}{Volker Markl}.}
  \bibinfo{year}{2022}\natexlab{}.
\newblock \showarticletitle{{Query Processing on Heterogeneous CPU/GPU
  Systems}}.
\newblock \bibinfo{journal}{\emph{ACM Comput. Surv.}} \bibinfo{volume}{55},
  \bibinfo{number}{1}, Article \bibinfo{articleno}{11} (\bibinfo{year}{2022}),
  \bibinfo{numpages}{38}~pages.
\newblock
\showISSN{0360-0300}
\urldef\tempurl%
\url{https://doi.org/10.1145/3485126}
\showDOI{\tempurl}


\bibitem[Sax(2018)]%
        {sax2018kafka}
\bibfield{author}{\bibinfo{person}{Matthias~J. Sax}.}
  \bibinfo{year}{2018}\natexlab{}.
\newblock \bibinfo{booktitle}{\emph{{Apache Kafka}}}.
\newblock \bibinfo{publisher}{Springer International Publishing},
  \bibinfo{address}{Cham}, \bibinfo{pages}{1--8}.
\newblock
\showISBNx{978-3-319-63962-8}
\urldef\tempurl%
\url{https://doi.org/10.1007/978-3-319-63962-8_196-1}
\showDOI{\tempurl}


\bibitem[Subramanian et~al\mbox{.}(2021)]%
        {subramanian2021gpu-db-operators-benchmark}
\bibfield{author}{\bibinfo{person}{Harish Kumar~Harihara Subramanian},
  \bibinfo{person}{Bala Gurumurthy}, \bibinfo{person}{Gabriel~Campero Durand},
  \bibinfo{person}{David Broneske}, {and} \bibinfo{person}{Gunter Saake}.}
  \bibinfo{year}{2021}\natexlab{}.
\newblock \showarticletitle{{Analysis of GPU-Libraries for Rapid Prototyping
  Database Operations: A look into library support for database operations}}.
\newblock \bibinfo{journal}{\emph{Proceedings of the 37th International
  Conference on Data Engineering Workshops (ICDEW)}}  \bibinfo{volume}{1}
  (\bibinfo{year}{2021}), \bibinfo{pages}{36--41}.
\newblock
\urldef\tempurl%
\url{https://doi.org/10.1109/ICDEW53142.2021.00014}
\showDOI{\tempurl}


\bibitem[Sun et~al\mbox{.}(2019)]%
        {sun2019bert4rec}
\bibfield{author}{\bibinfo{person}{Fei Sun}, \bibinfo{person}{Jun Liu},
  \bibinfo{person}{Jian Wu}, \bibinfo{person}{Changhua Pei},
  \bibinfo{person}{Xiao Lin}, \bibinfo{person}{Wenwu Ou}, {and}
  \bibinfo{person}{Peng Jiang}.} \bibinfo{year}{2019}\natexlab{}.
\newblock \showarticletitle{{BERT4Rec: Sequential Recommendation with
  Bidirectional Encoder Representations from Transformer}}. In
  \bibinfo{booktitle}{\emph{Proceedings of the 28th ACM International
  Conference on Information and Knowledge Management}} (Beijing, China)
  \emph{(\bibinfo{series}{CIKM '19})}. \bibinfo{publisher}{Association for
  Computing Machinery}, \bibinfo{address}{New York, NY, USA},
  \bibinfo{pages}{1441--1450}.
\newblock
\showISBNx{9781450369763}
\urldef\tempurl%
\url{https://doi.org/10.1145/3357384.3357895}
\showDOI{\tempurl}


\bibitem[Vaswani et~al\mbox{.}(2017)]%
        {vaswani2017attention}
\bibfield{author}{\bibinfo{person}{Ashish Vaswani}, \bibinfo{person}{Noam
  Shazeer}, \bibinfo{person}{Niki Parmar}, \bibinfo{person}{Jakob Uszkoreit},
  \bibinfo{person}{Llion Jones}, \bibinfo{person}{Aidan~N. Gomez},
  \bibinfo{person}{\L{}ukasz Kaiser}, {and} \bibinfo{person}{Illia
  Polosukhin}.} \bibinfo{year}{2017}\natexlab{}.
\newblock \showarticletitle{{Attention is All You Need}}. In
  \bibinfo{booktitle}{\emph{Proceedings of the 31st International Conference on
  Neural Information Processing Systems}} (Long Beach, California, USA)
  \emph{(\bibinfo{series}{NIPS'17})}. \bibinfo{publisher}{Curran Associates
  Inc.}, \bibinfo{address}{Red Hook, NY, USA}, \bibinfo{pages}{6000--6010}.
\newblock
\showISBNx{9781510860964}


\bibitem[Wilkening et~al\mbox{.}(2021)]%
        {RecSSD}
\bibfield{author}{\bibinfo{person}{Mark Wilkening}, \bibinfo{person}{Udit
  Gupta}, \bibinfo{person}{Samuel Hsia}, \bibinfo{person}{Caroline Trippel},
  \bibinfo{person}{Carole-Jean Wu}, \bibinfo{person}{David Brooks}, {and}
  \bibinfo{person}{Gu-Yeon Wei}.} \bibinfo{year}{2021}\natexlab{}.
\newblock \showarticletitle{RecSSD: Near Data Processing for Solid State Drive
  Based Recommendation Inference}. In \bibinfo{booktitle}{\emph{Proceedings of
  the 26th ACM International Conference on Architectural Support for
  Programming Languages and Operating Systems}} (Virtual, USA)
  \emph{(\bibinfo{series}{ASPLOS 2021})}. \bibinfo{publisher}{Association for
  Computing Machinery}, \bibinfo{address}{New York, NY, USA},
  \bibinfo{pages}{717--729}.
\newblock
\showISBNx{9781450383172}
\urldef\tempurl%
\url{https://doi.org/10.1145/3445814.3446763}
\showDOI{\tempurl}


\bibitem[Yuan et~al\mbox{.}(2013)]%
        {yuan2013gpu-query-processing}
\bibfield{author}{\bibinfo{person}{Yuan Yuan}, \bibinfo{person}{Rubao Lee},
  {and} \bibinfo{person}{Xiaodong Zhang}.} \bibinfo{year}{2013}\natexlab{}.
\newblock \showarticletitle{{The Yin and Yang of Processing Data Warehousing
  Queries on GPU Devices}}.
\newblock \bibinfo{journal}{\emph{Proc. VLDB Endow.}} \bibinfo{volume}{6},
  \bibinfo{number}{10} (\bibinfo{year}{2013}), \bibinfo{pages}{817--828}.
\newblock
\showISSN{2150-8097}
\urldef\tempurl%
\url{https://doi.org/10.14778/2536206.2536210}
\showDOI{\tempurl}


\bibitem[Zhao et~al\mbox{.}(2020)]%
        {Distributed_Hierarchical}
\bibfield{author}{\bibinfo{person}{Weijie Zhao}, \bibinfo{person}{Deping Xie},
  \bibinfo{person}{Ronglai Jia}, \bibinfo{person}{Yulei Qian},
  \bibinfo{person}{Ruiquan Ding}, \bibinfo{person}{Mingming Sun}, {and}
  \bibinfo{person}{Ping Li}.} \bibinfo{year}{2020}\natexlab{}.
\newblock \showarticletitle{{Distributed Hierarchical GPU Parameter Server for
  Massive Scale Deep Learning Ads Systems}}.
\newblock \bibinfo{journal}{\emph{Proceedings of Machine Learning and Systems}}
   \bibinfo{volume}{2} (\bibinfo{year}{2020}), \bibinfo{pages}{412--428}.
\newblock


\end{thebibliography}


\end{document}